\documentclass[a4paper,12pt]{article}

\usepackage{epsfig,graphicx,amsmath,amssymb}
\usepackage{url}
\usepackage{subcaption}
\usepackage[percent]{overpic}
\usepackage{siunitx}
\usepackage{units}
\usepackage{hyperref}
\begin{document}

\thispagestyle{empty}

$\phantom{.}$

\begin{flushright}
{\sf  MITP/14-110 \\
  } 
\end{flushright}

\hfill

\begin{center}
{\Large {\bf Mini-Proceedings, 16th meeting of the Working Group on Radiative Corrections and MC Generators for Low Energies} \\
\vspace{0.75cm}}

\vspace{1cm}

{\large 18$^{\mathrm{th}}$ - 19$^{\mathrm{th}}$ November, Laboratori Nazionali di Frascati, Italy}

\vspace{2cm}

{\it Editors}\\
Henryk Czyz (Katowice), Pere Masjuan (Mainz), Graziano Venanzoni (Frascati)
\vspace{2.5cm}

ABSTRACT

\end{center}

\vspace{0.3cm}

\noindent
The mini-proceedings of the 16$^{\mathrm{th}}$ Meeting of the "Working Group on Radiative Corrections and MonteCarlo Generators for Low Energies" held in Frascati, 18$^{\mathrm{th}}$ - 19$^{\mathrm{th}}$ November, are presented. These meetings, started in 2006, have as aim to bring together experimentalists and theoreticians working in the fields of meson transition form factors, hadronic contributions to the anomalous magnetic moment of the leptons, and the effective fine structure constant. The development of MonteCarlo generators and Radiative Corrections for precision $e^+e^-$ and $\tau$-lepton physics are also covered.

\medskip\noindent
The web page of the conference:
\begin{center}
\url{https://agenda.infn.it/conferenceDisplay.py?ovw=True&confId=8626} 
\end{center}
\noindent
contains the presentations.
\vspace{0.5cm}

\noindent
We acknowledge the support and hospitality of the Laboratori Nazionali di Frascati.

\newpage

{$\phantom{=}$}

\vspace{0.5cm}

\tableofcontents

\newpage

\section{Introduction to the Workshop}

\addtocontents{toc}{\hspace{1cm}{\sl H.~Czy\.z and G.~Venanzoni}\par}

\vspace{5mm}

\noindent
 H.~Czy\.z$^1$ and G.~Venanzoni$^2$

\vspace{5mm}

\noindent
$^1$Institute of Physics, University of Silesia, 40007 Katowice, Poland\\
$^2$Laboratori Nazionali di Frascati dellÕINFN, 00044 Frascati, Italy

\vspace{5mm}

The importance of continuous and close collaboration between the experimental
and theoretical groups is crucial in the quest for
precision in hadronic physics.
This is the reason why the  
Working Group on ``Radiative Corrections and Monte Carlo Generators for Low Energies'' (Radio MonteCarLow)  was formed a few years ago bringing together experts (theorists and experimentalists) working in the field of low-energy $e^+e^-$ physics and partly also the $\tau$ community.
Its main motivation  was to understand the status and the precision of the Monte Carlo generators used to analyze the hadronic cross section measurements
obtained as well with energy scans as with radiative return, to determine luminosities, and whatever possible to perform tuned comparisons, {\it i.e.}
comparisons of MC generators with a common set of input parameters and experimental cuts. This  main effort was summarized in a report published in 2010~\cite{Actis:2010gg}.
During the years the WG structure has been enriched of more physics items 
and now it includes seven subgroups: Luminosity, R-measurement, ISR, 
Hadronic VP $g-2$ and Delta alpha, gamma-gamma physics, FSR models, tau.

During the workshop the last achievements of each subgroups have been presented.
The present accuracy and the future prospects of MC generators for  $e^+e^-$ into leptonic, $\gamma\gamma$, and hadronic final states have been reviewed. 
The recent evaluation of the positronium contribution to the electron $g-2$ and the role of experimental data to the hadronic LO and Light-by-Light NLO contributions to the $g-2$ of the muon have been discussed.
New results from CMD3 and BESIII experiments have been presented.
Finally the status of HPrecisionNet work package of the networking program HPH to Horizon 2020, was presented.

The workshop was held from the 18$^{th}$ to the 19$^{th}$ November, at the Laboratori Nazionali di Frascati dellÕINFN, Italy.

Webpage of the conference is 
\begin{center}
\url{https://agenda.infn.it/conferenceDisplay.py?ovw=True&confId=8626} 
\end{center}
\noindent
where detailed program and talks can be found.

All the information on the WG can be found at the web page:
\begin{center}
\url{http://www.lnf.infn.it/wg/sighad/} 
\end{center}

\newpage

\section{Summaries of the talks}

\subsection{Present accuracy and future prospects of
Monte Carlo generators for Bhabha and $e^+e^-\to\gamma\gamma$}
\addtocontents{toc}{\hspace{2cm}{\sl C.M. Carloni}\par}

\vspace{5mm}

C.M. Carloni Calame

\vspace{5mm}

\noindent
Dipartimento di Fisica, Universit\`a di Pavia, Via A. Bassi 6, 
27100 Pavia, Italy
\vspace{5mm}

The
knowledge of the luminosity $\mathcal L$ is a key ingredient
for any measurement at $e^+e^-$ machines. The
usual
strategy to
calculate it is through the relation ${\mathcal L} = N/\sigma_{th}$,
where $\sigma_{th}$ is the theoretical cross section of a
QED process and $N$ the number of events. QED
processes are the best choice because of their clean signal, low background and the
possibility to push the theoretical accuracy up to the $0.1\%$
level or better. The latter requires the inclusion of the relevant radiative
corrections (RCs) and their implementation into Monte Carlo (MC) event
generators (EGs) to easily account for experimental event selection criteria.

Modern EGs used for luminometry simulate Bhabha,
$e^+e^-\to\gamma\gamma$ and $e^+e^-\to\mu^+\mu^-$ (or a
sub-set of them), including the exact NLO QED corrections and/or a
leading-log (LL) approximation of
higher-order (h.o.)
effects~\cite{babayaga,bhwide,mcjpg,dragovenanzoni,BK}. 
The consistent inclusion of NLO and h.o. LL contributions is mandatory in
view of the required theoretical accuracy.

Focusing on Bhabha scattering, in order to estimate the theoretical
error of the EGs, it is
extremely important to perform tuned comparisons among them, to assess
the technical precision and have an idea of the accuracy of
the included corrections, usually implemented according to
different approaches. This has been done and reported
in~\cite{quest}, from which tab.~\ref{tab1} has been extracted.
\begin{table}[h]
$$
  \begin{array}{|c|c|c|c|c|}
    \hline
    {\rm setup} &{\tt BabaYaga@NLO}&{\tt BHWIDE}&{\tt MCGPJ}&{\delta (\%)}\\[1mm]
    \hline
\sqrt{s}=1.02~{\rm GeV}, 20^{\circ} \leq \vartheta_{\mp} \leq 160^{\circ} & 6086.6(1) & 6086.3(2)& $---$ &{0.005}\\[1mm]
{\sqrt{s}=1.02~{\rm GeV}, 55^{\circ} \leq \vartheta_{\mp} \leq 125^{\circ} } & 455.85(1) & 455.73(1)& $---$ &{0.030}\\[1mm]
{\sqrt{s}= 3.5~{\rm GeV}, |\vartheta_+ + \vartheta_- - \pi| \leq 0.25~{\rm rad}} & 35.20(2) & $---$ & 35.181(5) &{0.050}\\[1mm]
\sqrt{s}= 10~{\rm GeV},  40^{\circ} \leq \vartheta_{\mp} \leq
  160^{\circ} & 11.67(3) & 11.660(8) & $---$ & 0.086\\[1mm]
    \hline    
  \end{array}
$$
\vspace{-3mm}
\caption{Comparison of Bhabha cross sections (in nb) for
different setups, obtained with the EGs described
in~\cite{babayaga},~\cite{bhwide} and~\cite{mcjpg}.
See~\cite{quest} for further details and results.}
\label{tab1}
\end{table}
In general, it is found that the different MC EGs predict cross
sections which differ by at most $0.1\%$ when including NLO and
h.o. LL corrections. This is in fair agreement with the accuracy
of the different approaches estimated by the authors. 

A further step to put on firmer ground the theoretical error is to
compare with exact NNLO results, which have been calculated for
Bhabha scattering by various groups in the last years (see references
in~\cite{quest}). NNLO RCs are
partly included in EGs and, once extracted, their NNLO contributions
can be consistently compared with exact calculations.
This has been done for example in the $3^{\rm rd}$ paper
of~\cite{babayaga} and in~\cite{heavypairs}.

In Tab.~\ref{tab2}
(adapted from~\cite{quest}) the
total error ``budget'' is summarized for typical luminometry
conditions at flavor factories. The main conclusion to draw from
it is that RCs currently implemented into MC EGs allow to reach
a theoretical precision up to the $0.1\%$ level.
\begin{table}[thb]
\begin{center}
\begin{tabular}{llll}
\hline
{ Source of error (\%)} & { $\Phi-$factories}  & 
{\scriptsize $\sqrt{s}$ = 3.5 GeV} & { $B-$factories} \\
\hline
$|\delta^{\rm err}_{\rm VP}|$      & 0.02 & 0.01 & 0.02 \\ 
\hline
\hline
 $|\delta^{\rm err}_{{\rm SV},\alpha^2}|$ & 0.02 & 0.02 & 0.02\\
 \hline
 $|\delta^{\rm err}_{{\rm HH},\alpha^2}|$       & 0.00 & 0.00 & 0.00\\
 \hline
 $|\delta^{\rm err}_{{\rm SV,H},\alpha^2}|$    & 0.05 & 0.05 & 0.05 \\
  \hline
 $|\delta^{\rm err}_{\rm pairs}|$  & 0.03 & 0.016  & 0.03
  \\
\hline
\hline
 { $|\delta^{\rm err}_{\rm total}|$ linearly}   & { 0.12} & { 0.1} &  { 0.13} \\
 \hline
  { $|\delta^{\rm err}_{\rm total}|$ in quadrature}   & { 0.07} & 
  { 0.06} &  { 0.06} \\
\hline
\end{tabular}
\end{center}
\vspace{-3mm}
\caption{Total error ``budget'' for Bhabha cross section at flavor
  factories. See~\cite{babayaga,quest,heavypairs} for more details and
  definitions.}
\label{tab2}
\end{table}

It has to be mentioned that the error induced by vacuum polarization (VP)
corrections is driven and dominated by experimental errors. At
energies around the narrow resonances (such as $J/\Psi$), VP errors
might be larger than in tab.~\ref{tab2} and a dedicated study is needed.

A similar picture ought to be valid also for $\gamma\gamma$ final state, with
the added advantage that, at least up to NLO, VP corrections do not
contribute to the cross section. Nevertheless, a careful estimate of
the theoretical error in this case has not been performed yet and would be of high interest.

A possible improvement of the theoretical accuracy, if going beyond
the $0.1\%$ level is needed at all, would be the inclusion of the full
NNLO results into the MC EGs, which is a non trivial but feasible task.

\vspace{2mm}
\noindent I'd like to thank H.~Czyz and G.~Venanzoni for the kind invitation
and the organization of a really stimulating workshop.

\newpage

\subsection{Current status of luminosity measurement with the CMD-3 detector at the VEPP-2000 e$^+$e$^-$ collider}
\addtocontents{toc}{\hspace{2cm}{\sl G.V.Fedotovich}\par}

\vspace{5mm}

G.V.Fedotovich$^{1,2}$, A.E.Ryzhenenkov$^{1,2}$

\vspace{5mm}

\noindent
$^1$Budker Institute of Nuclear Physics, SB RAS, Novosibirsk, 630090, Russia\\
$^2$Novosibirsk State University, Novosibirsk, 630090, Russia
\vspace{5mm}

Since December 2010 the CMD-3~\cite{Khazin:2008} detector has taken data at the electron-positron collider VEPP-2000~\cite{Berkaev:2012}. The collected data sample 
corresponds to an integrated luminosity of 60 pb$^{-1}$ in the c.m. energy 
from 0.32 up to 2~GeV.

The luminosity is a key part in many experiments which study the hadronic 
cross sections at $e^+e^-$ colliders. As a rule, the systematic 
error of the luminosity determination represents one of the largest sources of uncertainty 
which can cause significant reduction of the hadronic cross sections accuracy.
Therefore it is very important to have several well known QED 
processes such as $e^+e^- \to e^+e^-$, $\mu^+ \mu^-$, $\gamma\gamma$ to determine 
the luminosity. The combined application of them will help to better understand and 
estimate a real systematic accuracy of the luminosity. The combined application of them will
help to better understand and estimate a real systematic accuracy of the luminosity. 

The energy range from 1 to 2 GeV was scanned up and down with a step of 50 MeV. 
At each energy point the integrated luminosity $\sim$ 500 nb$^{-1}$ was collected. 
During the scan down the energy points, at which the data were collected, have been shifted 
to the previous one by 25 MeV. The data were collected at an average luminosity 
$\sim4\cdot10^{30}$ s$^{-1}\cdot$cm$^{-2}$. At the highest energies 
the peak luminosity reached the values about $2\cdot10^{31}$ s$^{-1}\cdot$cm$^{-2}$
and was restricted by the positron storage rate in the booster. The project luminosity 
$\sim10^{32}$ s$^{-1}\cdot$cm$^{-2}$ will be provided only with start of operating of new positron injection facility in 2015.
The beam energy has been monitored 
($\sim$ 0.5 MeV) by measuring the current in dipole magnets of the main ring. 
The time period of this run 
was extend from January to June 2011. 
In 2012 the luminosity was measured at 16 energy points from 1.32~GeV to 1.98~GeV and collected 
luminosity was about $\sim$ 14 pb$^{-1}$. 

In 2013 the energy range from 0.32~GeV to 1~GeV was scan with the 10~MeV step. The integrated 
luminosities about 8.3 and 8.4 pb$^{-1}$ were collected around $\omega$ and $\phi$ mesons. Over the 
2013 year the integrated luminosity $\sim$ 25~pb$^{-1}$ has been collected.

\begin{figure}[th]
\begin{minipage}[t]{0.46\textwidth}
\centerline{\includegraphics[width=0.75\textwidth]{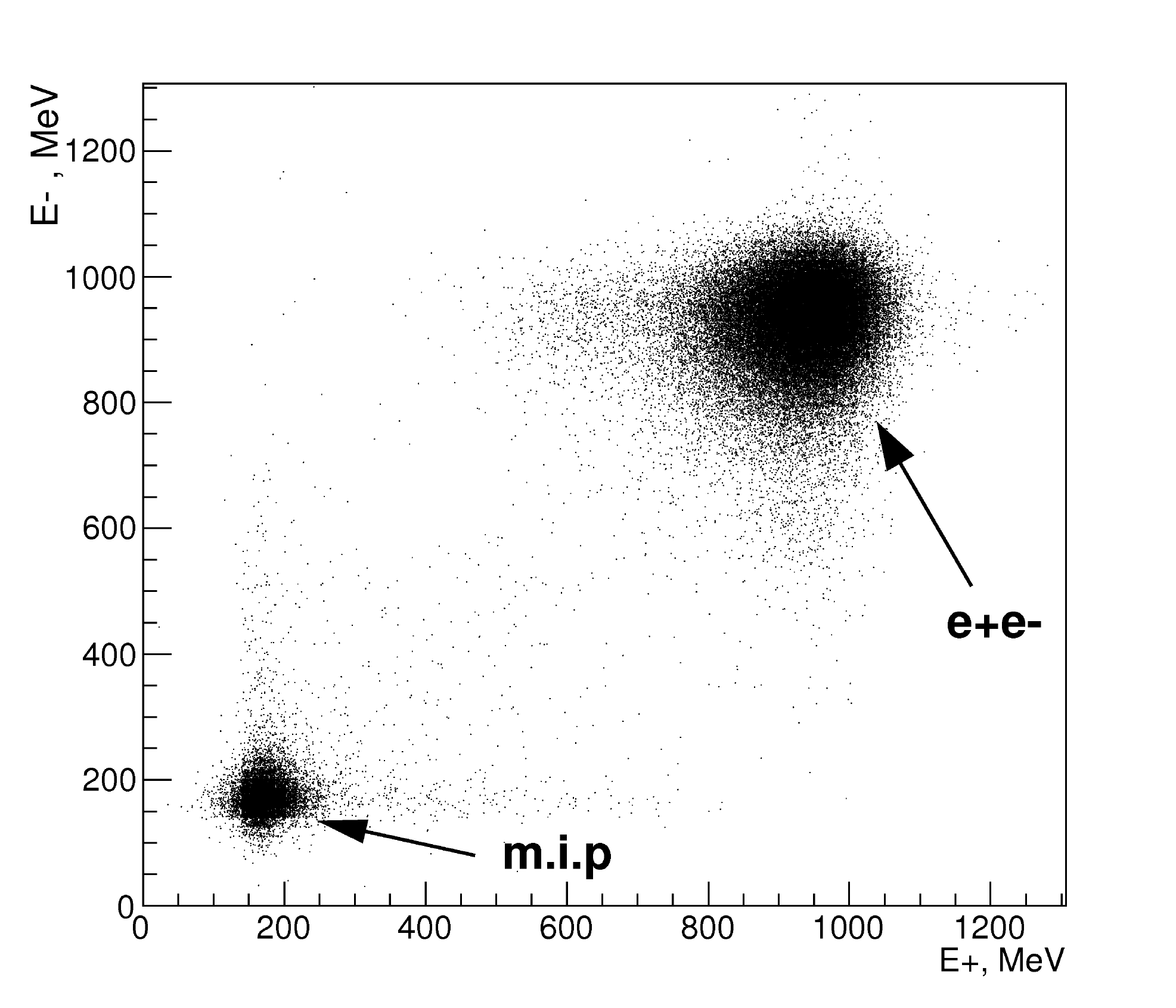}}
\caption{Two-dimensional plot of energy deposition in calorimeters one particle vs 
another for collinear tracks.}
\label{e1vse2}
\end{minipage}\hfill\hfill
\begin{minipage}[t]{0.46\textwidth}
\centerline{\includegraphics[width=0.98\textwidth]{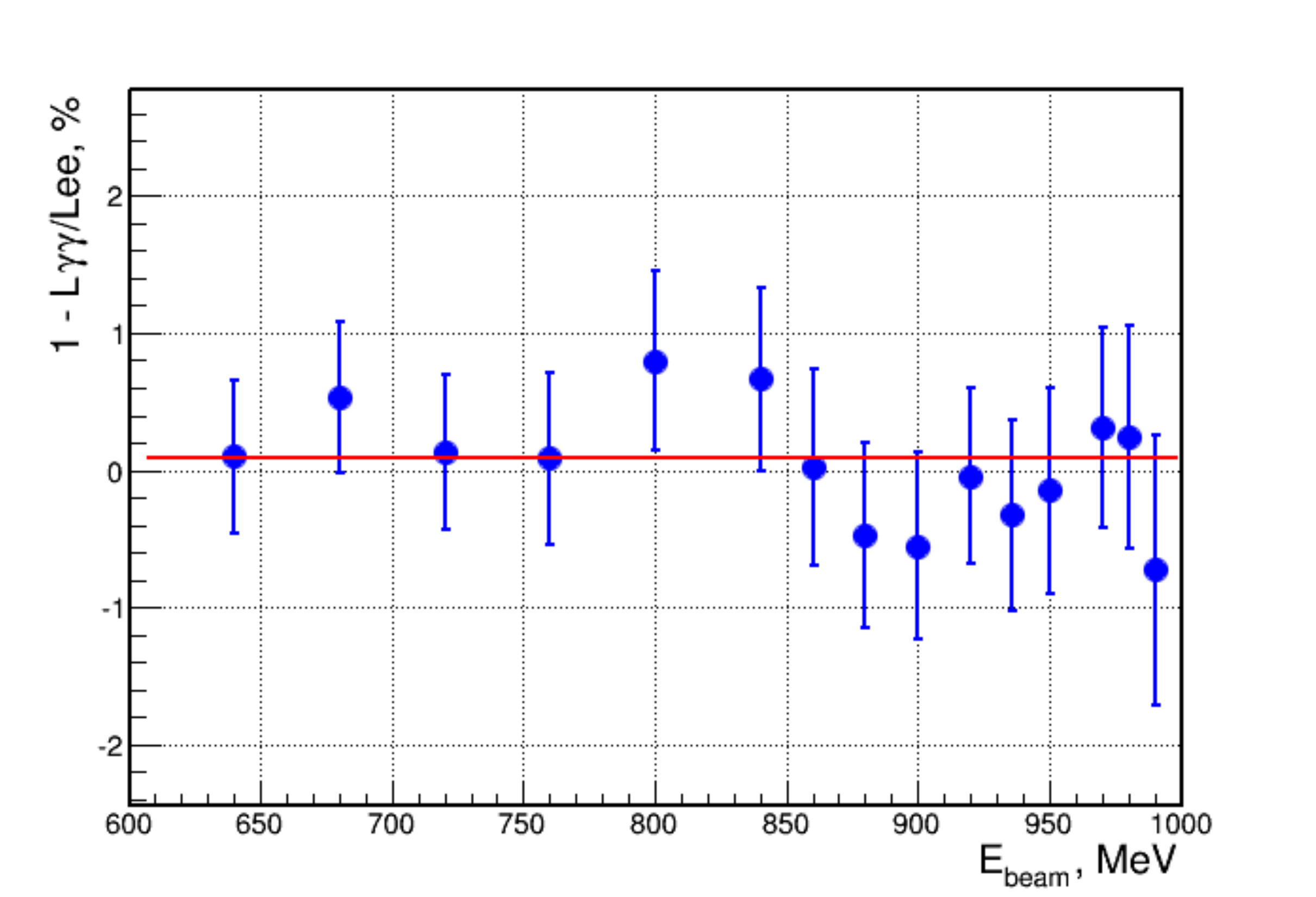}}
\caption{The ratio of the luminosities for the $e^+e^-$ and $\gamma\gamma$ processes vs energy. Scan 2012. 
The horizontal red line - fit.}
\label{scan12}
\end{minipage}
\end{figure}

The sample of collinear events $e^+e^-$, $\mu^+\mu^-$, $\pi^+\pi^-$, $K^+K^-$ and 
cosmic background were selected for luminosity determination. The two-dimensional plot of energy deposition 
in calorimeters for these events is presented in 
Fig.\ref{e1vse2} 
for the beam energy 950 MeV. It is clear seen that Bhabha events are distributed predominantly at the 
upper right corner whereas other particles are concentrated in the bottom left one. Thus, the integrated 
luminosity can be determined by the well selected Bhabha events.

To select $\gamma\gamma$ events the information about they energy deposition 
in calorimeters is used. 
It is seen that the signal events are concentrated as a cluster of dots in upper-right corner of this plot. 
At the same time two train seen - concentration of dots in two mutually perpendicular directions due to ISR.
The events of this sample should have energy deposition inside interval: 
$0.5E_{beam } < E_{0}, E_{1} < 1.5E_{beam}$. Unfortunately the small part of the 
Bhabha events can seep under the central peak and imitate $\gamma\gamma$ events. To exclude such events the additional 
condition was applied - the Z-chamber sectors associated with clusters must be triggered. Visual scan of the 
remaining events proofed - there are not Bhabha events under central. Unfortunately 
this condition delete some $\gamma\gamma$ events due to albedo coming from showers. The fraction of such events 
amounts to $\sim 6\%$ and as a result we should include correction about $0.36\%$ to restore the number 
of $\gamma\gamma$ events.
    
\begin{figure}[tbh]
\end{figure}
\begin{figure}[tbh]
\begin{minipage}[t]{0.46\textwidth}
\centerline{\includegraphics[width=0.98\textwidth]{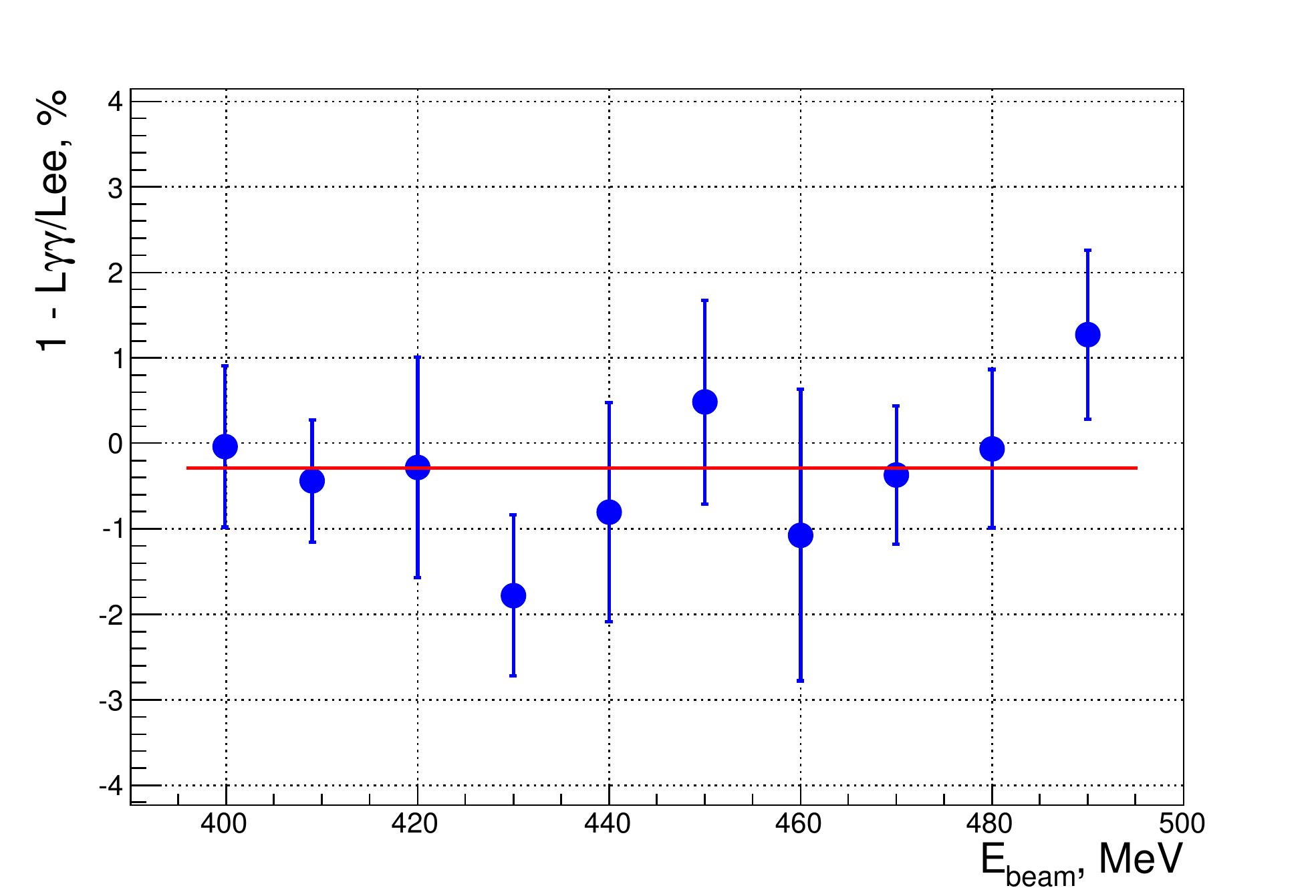}}
\caption{The ratio of the luminosities for the $e^+e^-$ and $\gamma\gamma$ processes vs energy. Scan 2013. 
The horizontal red line - fit.}
\label{scan13}
\end{minipage}\hfill\hfill
\begin{minipage}[t]{0.46\textwidth}
\centerline{\includegraphics[width=0.98\textwidth]{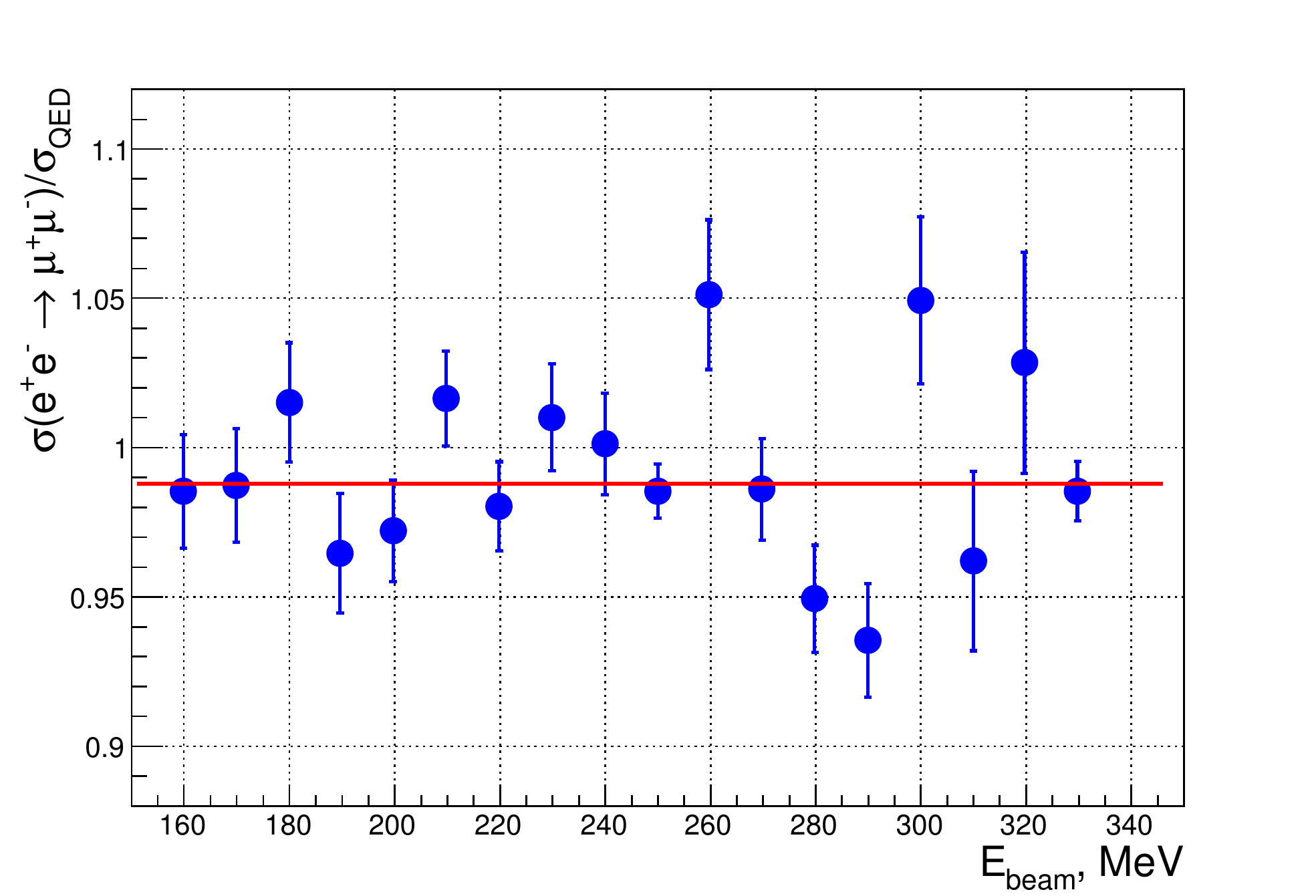}}
\caption{The results of cross section measurement of muon production in comparison with the prediction of QED.}
\label{nmmnee}
\end{minipage}
\end{figure}    
    
The luminosities ratio determined with use of two processes vs energy is presented in 
Fig.~\ref{scan12} and in Fig.~\ref{scan13}, 
where only statistical errors are shown. The blue circles correspond to the scan up, whereas red circles - scan 
down. The horizontal line is a fit for this ratio for scan down. In this case the relative difference between 
luminosities is in average $(0.73 \pm 0.35)$\%. However, at the beginning of the run the difference was
$\sim$3\% and explained by hardware problems and the quality of 
inter-calibration of the detector subsystems. Collecting all the main sources which contribute to systematic error of the luminosity,
we estimate the current accuracy as $\sim$1\% while. The first energy scan below 1~GeV was performed at VEPP-2000 during the season of 2013. 
The preliminary results of the luminosity measurement are shown in
Fig.~\ref{scan13}.
The already collected statistics is higher than that in the previous CMD-2 experiment and at the level or 
better than in BaBar and KLOE experiments. One of the tests in this analysis is to measure the cross 
section of the process $e^+e^- \to \mu^+\mu^-$ at low energy, where particles separation is possible using 
only momentum information from DC. Preliminary results of this test are consistent with the QED prediction 
as it is seen in 
Fig.~\ref{nmmnee}. 
The radiative corrections (RC) to this cross section with photon jets radiation 
in collinear regions were taken into account according to
\cite{mcgpj} and their accuracy is better than $0.2\%$.

\newpage

\subsection{The role of experimental data as input information for precise hadronic calculations: muon $g-2$, rare $\pi^0$ decays, and mixing parameters }
\addtocontents{toc}{\hspace{2cm}{\sl P.~Masjuan}\par}

\vspace{5mm}

P.~Masjuan

\vspace{5mm}

\noindent
PRISMA Cluster of Excellence, Institut f\"ur Kernphysik, Johannes Gutenberg-Universt\"at, Mainz D-55099, Germany\\

\vspace{5mm}

One of the open questions concerning the Hadronic Light-by-Light scattering contribution to the muon $g-2$ (HLBL) is the role of experimental data.

Part of the difficulty of including experimental data in the HLBL is due to the particular framework where the main calculations are done~\cite{Jegerlehner:2009ry}, the large-$N_c$ of QCD~\cite{tHooft:1973jz}. In such limit, one uses the resonance saturation scheme to reproduce the pseudoscalar transition form factor (TFF) that appears in the dominant piece of the HLBL, the pseudscalar-exchange contribution~\cite{Jegerlehner:2009ry}. The main inputs are, then, the pion decay constant and the values of the resonance masses. On top, even though data on the TFF are willing to be included, one still faces the problem on how to link the different kinematic regimes between the experiment for the TFF (low-energy region at time- and space-like, together with intermediate energies at space like)  and the kinematics for the pseudoscalar-exchange diagram (whole space-like energy region, from origin of energies up to infinity)~\cite{Jegerlehner:2009ry}. 

In this talk I summarized our attempt to provide an answer to that question in a model-independent fashion~\cite{ourpaper}\footnote{For a recent summary, see for instance Ref.~\cite{Masjuan:2014rea}}, a method based on the analyticity of the TFF, compatible with the recent dispersion relations approach~\cite{mh} with the advantage of having larger photon energy range of applicability (in practice, the full energy range), and based on the low-energy properties of the TFF. This endeavor started two years ago in Ref.~\cite{Masjuan:2012wy} and was further developed in Refs.~\cite{ourpaper,Masjuan:2014rea,Escribano:2013kba}.

The method proposed can be summarized as follows:
\begin{itemize}
\item our attempt is, indeed, a method, not \textit{a model}. 
\item it shall be simple, easy to understand, to apply and reproduce, in contrast to more involved procedures such as dispersion relations.
\item it may contain approaches (to say, improvable), but not assumptions (not improvable).
\item it should be systematic, easy to update with new experimental data but also it should provide a systematic error, a pure error from the method itself.
\item finally, it should be predictive meaning checkable
\end{itemize}

The method proposed is based on the mathematical theory of Pad\'e approximants (PA). It was pointed out in Ref.~\cite{Masjuan:2007ay} that, in the large-$N_c$ framework, the resonance saturation scheme employed in the HLBL can be understood from the theory of PA to meromorphic functions~\cite{Baker}, where one can compute the desired quantities in a model-independent way. Also, the analytical properties of the TFF indicate that the convergence of the PA is guaranteed at the energies we are working with. Altogether defines the systematics of our method (more iterations should give better approximation) and ascribe an error to that~\cite{Masjuan:2008cp}. 

To exemplify the advantages of our method, we considered a model from Ref.~\cite{Knecht:2001qf}. As stated before, the inputs for the model can have two different sources: first, a pure theoretical origin based on large-$N_c$ and chiral limits (inputs are resonance masses within the \textit{half-width rule}~\cite{Masjuan:2012gc} and the meson decay constant in the chiral $SU(3)$ limit~\cite{Ecker}); second, a reconstruction of the models based on a matching with the TFF low-energy constants~\cite{Masjuan:2012wy}, i.e, \textit{\'a la} PA~\cite{Masjuan:2007ay,Baker,Masjuan:2008cp} minimizing in such a way the model dependence~(see \cite{Masjuan:2014rea} for details). The former yields a final error for the $\pi^0$ contribution to HLBL to $15\%$ ($5\%$ from $F_0$ and $10\%$ from the masses). The later, the Pad\'e method, yield a similar $15\%$  provided that the $13\%$ error on the slope ($25\%$ on curvature) implies an error of $10\%$ ($5\%$) in the pion contribution; the impact of $F_{\pi}$ is dramatic since its $1\%$ error implies a $2\%$ error on HLBL. Interestingly enough, the central value is $20\%$ higher, driving non compatible results! The PA method, predictable~\cite{Aguar-Bartolome:2013vpw}, can accommodate space- and time-like data. Provides also a rule-of-thumb for estimating the impact of experimental uncertainties, a point never discussed before.

Notice, nevertheless, that the standard procedure~\cite{Jegerlehner:2009ry,Knecht:2001qf} to treat the TFF  is through a factorization approach, e.g.,  $F(Q_1^2,Q_2^2)=F(Q_1^2,0)\times F(0,Q_2^2)$ where $F(Q^2,0)$ is the measured quantity. The impact of such approach is not negligible (see my contribution in~\cite{Adlarson:2014hka}).

In conclusion, we remark the important role of experimental data to determine the dominant pieces of the HLBL (i.e., $\pi^0,\eta,\eta'$). We argue that the way of including such information should be based on PA which provides a systematic error and a simple rule for estimating the impact of experimental uncertainties, both from the space- and the time-like~\cite{ourpaper}. We notice, finally, that the errors discussed above have been unfortunately ignored in the main reviews (no error for $F_0$ or resonance masses have been properly estimated, neither the possibility to match with experimental low-energy description of the TFF) and that posses a warrant on the reliability of the current error estimates for the HLBL. Similar discussions concerning $P\to l^+l^-$ as well as mixing angles were also addressed.

As discussed during the meeting, to extract resonance poles using PA, see~Ref.\cite{poles}.

\newpage

\subsection{On the positronium contribution to the electron $g-2$}
\addtocontents{toc}{\hspace{2cm}{\sl M.~Passera}\par}

\vspace{5mm}

M.~Fael$^{1}$ and M.~Passera$^{2}$

\vspace{5mm}

\noindent
$^{1}$~Albert Einstein Center for Fundamental Physics, Institute for Theoretical Physics, University of Bern, CH-3012 Bern, Switzerland\\
$^{2}$~Istituto Nazionale Fisica Nucleare, Sezione di Padova, I-35131 Padova, Italy

\vspace{5mm}

The leading contribution of positronium, the $e^+$$e^-$ bound state, to the anomalous magnetic moment of the electron $(a_e)$ has been computed in Ref.~\cite{Mishima:2013ama}. The result of this calculation,
\begin{equation}
	a_e^{\rm  P} = \frac{\alpha^5}{4\pi} \zeta(3) \left( 8\ln2 - \frac{11}{2} \right) =  0.89 \times 10^{-13},
\label{eq:aeP}
\end{equation}
where $\zeta(3) = 1.202\ldots$ and $\alpha$ is the fine-structure constant, is of the same order of $\alpha$ as the perturbative QED five-loop contribution $a_e^{(10)} = 9.16 \, (58) \left( \alpha/\pi \right)^5$~\cite{Aoyama:2012wj} and comparable with the present experimental uncertainty $\delta a_e = 2.8 \times 10^{-13}$~\cite{Gabrielse}. As it seems reasonable to expect a reduction of $\delta a_e$ to a part in $10^{-13}$ (or better) in ongoing efforts to improve this measurement, and work is in progress to reduce the error induced in the theoretical prediction for $a_e$ by the uncertainty of $\alpha$~\cite{Bouchendira:2010es,Terranova:2013vfa}, a test of the electron $g$-2 at the level of $10^{-13}$ (or below) is a goal that may be achieved not too far in the future. This will bring $a_e$ to play a pivotal role in probing new physics and allow to test whether the long-standing 3--4$\sigma$ discrepancy in the muon $g$-2 also manifests itself in the electron one~\cite{Giudice:2012ms}.

Recently the authors of Ref.~\cite{Melnikov:2014lwa} pointed out the presence of the continuum nonperturbative contribution 
\begin{equation}
	a_e({\rm vp})^{\rm cont, np} = - \frac{|\alpha|^5}{8\pi} \zeta(3) \left( 8\ln2 - \frac{11}{2} \right)
\label{eq:aeC}
\end{equation}
arising from the region right above the $s=4m^2$ threshold, which corresponds to $e^+ e^-$ scattering states with the exchange of Coulomb photons. Comparing Eqs.~(\ref{eq:aeP}) and (\ref{eq:aeC}) they showed that this additional ${\cal O}(\alpha^5)$ nonperturbative contribution cancels one-half of that of the positronium poles. The question is therefore how to deal with the remaining half: should one add it to the perturbative five-loop QED result of Ref.~\cite{Aoyama:2012wj}? Reference~\cite{Melnikov:2014lwa} argued that this remaining $a_e^{\rm  P}/2$ term is already contained in the perturbative ${\cal O}(\alpha^5)$ contribution to $a_e$ computed in Ref.~\cite{Aoyama:2012wj} and, therefore, it should not be added to it. On the other hand, in Ref.~\cite{Hayakawa:2014tla} it was claimed that positronium contributes to $a_e$ only through diagrams of ${\cal O}(\alpha^7)$ or higher. Also, on more general grounds~\cite{BraunM.A.:1968xia}, Ref.~\cite{Eides:2014swa} argued that $a_e^{\rm  P}$ does not exist.

In order to clarify this issue, in Ref.~\cite{Fael:2014nha} we used the closed form for the QED vacuum polarization function near the $s=4m^2$ threshold of Refs.~\cite{BraunM.A.:1968xia,Barbieri:1973lza} to verify that the total (positronium poles plus continuum) nonperturbative contribution to $a_e$ arising from the threshold region is equal to $a_e^{\rm  P}/2$. Then, using the analytic QED vacuum polarization at four-loop of Ref.~\cite{Baikov:2013ula}, we showed explicitly that the perturbative five-loop calculation of $a_e$ of Ref.~\cite{Aoyama:2012wj} does indeed contain the remaining term $a_e^{\rm  P}/2$, in agreement with the arguments of Ref.~\cite{Melnikov:2014lwa}. We also showed that this term $a_e^{\rm  P}/2$ arises from the class I(i) of five-loop diagrams of Ref.~\cite{Aoyama:2010zp} containing only one closed electron loop.

In conclusion, we showed explicitly that there is no additional contribution of QED bound states to $a_e$ beyond perturbation theory.

\vspace{2mm}{\it Acknowledgments} The work of M.F.\ is supported by the Swiss National Science Foundation. M.P.\  thanks the Department of Physics and Astronomy of the University of Padova for its support. His work was supported in part by the PRIN 2010-11 of the Italian MIUR and by the European Program INVISIBLES (PITN-GA-2011-289442).


\newpage

\subsection{Hadronic light-by-light scattering in the muon $g-2$: a dispersive approach}
\addtocontents{toc}{\hspace{2cm}{\sl M.~Hoferichter}\par}

\vspace{5mm}

M.~Hoferichter$^{1,2}$

\noindent
$^1$ Institut f\"ur Kernphysik, Technische Universit\"at
Darmstadt, Germany\\
$^2$ ExtreMe Matter Institute EMMI, GSI Helmholtzzentrum f\"ur
Schwerionenforschung GmbH, Germany\\
\vspace{5mm}

The uncertainty in the Standard-Model prediction for the anomalous magnetic moment of the muon is dominated by strong interactions. While hadronic vacuum polarization is intimately related to $e^+e^-\to\text{hadrons}$ by means of a dispersion integral, a similarly data-driven approach has only recently been suggested for hadronic light-by-light scattering (HLbL) (see~\cite{Kurz:2014wya,Colangelo:2014qya} for even higher-order hadronic contributions). Our framework~\cite{Colangelo:2014dfa,Colangelo:2014pva} exploits the analytic structure of the HLbL tensor, concentrating on pseudoscalar poles and two-meson intermediate states, which dominate at low energies.\footnote{A different approach, which aims at a dispersive description of the muon vertex function instead of the HLbL tensor, has
recently been presented in~\cite{Pauk:2014rfa}. An alternative strategy to reduce the model dependence in HLbL is based on lattice QCD~\cite{Blum:2014oka}.} 
The key input quantities
for such a program are the doubly-virtual pion transition form factor~\cite{Hoferichter:2014vra}
and the partial waves for $\gamma^*\gamma^*\to\pi\pi$~\cite{GM,Hoferichter:2011wk,Moussallam13,Hoferichter:2013ama,Dai:2014zta}, which in the absence of doubly-virtual data can again be reconstructed dispersively (see~\cite{Hanhart_eta} for a similar approach to the $\eta$, $\eta'$ transition form factor).

The calculation of the pion transition form factor~\cite{Hoferichter:2014vra} can be understood as a generalization of  existing analyses for $\gamma\pi\to\pi\pi$~\cite{Hoferichter:2012pm} and $\omega,\phi\to 3\pi$~\cite{Niecknig:2012sj}, which provide access to the form factor at fixed isoscalar virtualities $q_s^2=0$ and $ q_s^2=M_\omega^2,M_\phi^2$~\cite{Schneider:2012ez}, respectively. However, the normalization of the amplitude for a given $q_s^2$ cannot be predicted, but needs to be inferred from experiment, in case of $\gamma\pi\to\pi\pi$ by means of the Wess--Zumino--Witten anomaly, in case of $\omega,\phi\to 3\pi$ via the decay width. For general $q_s^2$, the normalization is extracted from $e^+e^-\to3\pi$ cross-section data, providing a prediction for $e^+e^-\to\pi^0\gamma$ without adjusting further parameters. So far, the phenomenological analysis of the singly-virtual form factor has been carried out, including accurate predictions for the slope of the form factor and its analytic continuation into the space-like region.  

A crucial step in the derivation of our dispersive formalism~\cite{Colangelo:2014dfa} concerns the construction of a suitable basis for the HLbL tensor, in such a way that the coefficient functions are free of kinematic singularities~\cite{Bardeen:1969aw,Tarrach:1975tu,Leo:1975fb}. 
Moreover, contributions involving double-spectral regions need to be considered separately, so that
the sQED pion loop augmented with pion vector form factors (FsQED), as identified on the level of the Mandelstam representation, is evaluated based on Feynman loop integrals.
In the talk, also a first numerical evaluation of $S$-wave $\pi\pi$ intermediate states was presented. Despite the double-spectral regions and solely based on $S$-waves the FsQED contribution can be reproduced at the $(5\text{--}10)\%$ level. Including $\pi\pi$ rescattering in the $\gamma^*\gamma^*\to\pi\pi$ partial waves in a simplified formalism that involves a Born-term left-hand cut and a finite matching point below the $K\bar K$ threshold, we find that the sum of $I=0$ and $I=2$ rescattering contributes $\sim-5\times 10^{-11}$ and, taken together with FsQED, $\sim-20\times 10^{-11}$ to HLbL scattering in the muon $g-2$.

\newpage

\subsection{Primary Monte-Carlo generator of the process $e^+e^-\to a_0(980)\rho(770)$ for the CMD-3 experiment }
\addtocontents{toc}{\hspace{2cm}{\sl P.A.~Lukin}\par}

\vspace{5mm}

P.A.~Lukin

\vspace{5mm}

\noindent
Budker Institute of Nuclear Physics and Novosibirsk State University, \\ Russia, 630090, Novosibirsk  \\

\vspace{5mm}

Electron-positron collider VEPP-2000~\cite{vepp} has been operating in Budker Institute of Nuclear Physics since 2010. 
Center-of-mass energy ($E_{c.m.}$) range covered by the collider is from threshold of hadron production and up to 2 GeV. 
Special optics, so called ``round beams'', used in the collider construction, allowed to obtain luminosity 
$2\times10^{31}$ cm$^{-2}\cdot$s$^{-1}$ at $E_{c.m.} = $1.8 GeV.   

The general purpose detector CMD-3 has been described in 
detail elsewhere~\cite{sndcmd3}. Its tracking system consists of a 
cylindrical drift chamber (DC)~\cite{dc} and double-layer multiwire 
proportional 
Z-chamber, both also used for a trigger, and both inside a thin 
(0.2~X$_0$) superconducting solenoid with a field of 1.3~T.
The liquid xenon (LXe) barrel calorimeter with 5.4~X$_0$ thickness has
fine electrode structure, providing 1-2 mm spatial resolution~\cite{lxe}, and
shares the cryostat vacuum volume with the superconducting solenoid.     
The barrel CsI crystal calorimeter with thickness 
of 8.1~X$_0$ is placed
outside  the LXe calorimeter,  and the end-cap BGO calorimeter with a 
thickness of 13.4~X$_0$ is placed inside the solenoid~\cite{cal}.
The luminosity is measured using events of Bhabha scattering 
at large angles~\cite{lum}. 

Physics program of the CMD-3 experiment includes the study of the multi-hadron production.
The cross section measurement of the $e^+e^-\to 3(\pi^+\pi^-)$ process in $E_{c.m.} = 1.5~--~2.0$ GeV has been already 
published~\cite{6pi}. §± Preliminary results the study 2$(\pi^+\pi^-\pi^0)$ final state has been reported~\cite{4pi2pi0}.

The study of intermediate states which lead to 2$(\pi^+\pi^-\pi^0)$ final state is essential to correctly describe the 
angular correlations between the particles and determine  the registration efficiency of the process under study. 
As it was reported at~\cite{MCWG_Apr14} the intermediate states $\omega(782)3\pi$, $\omega(782)\eta(545)$ and 
$\rho(770)(4\pi)_{S-wave}$ allow satisfactorily describe mass and angular distributions of  the 2$(\pi^+\pi^-\pi^0)$ 
production in $E_{c.m.} = 1.5~--~1.7$ GeV. 

But for higher $E_{c.m.}$ it is not possible to describe $\eta(545)$ signal, seen in three-pion mass distribution of 
the 2$(\pi^+\pi^-\pi^0)$, by contribution either $\omega(782)\eta(545)$ or $\phi(1020)\eta(545)$, because corresponding
cross sections are small enough. We supposed that $\eta(545)$ can be explained by the process 
$e^+e^-\to a_0(980)\rho(770)$  with dominant decay of $a_0(980)$ into $\eta(545)\pi$. The primary Monte-Carlo generator
for the process has been created out and implemented into the CMD-3 experiment Monte-Carlo simulation package. Using
the generator the signal of the $e^+e^-\to a_0(980)\rho(770)$ process has been observed in the experimental data for
2$(\pi^+\pi^-\pi^0)$ final state. 

However, it was not possible to describe 2$\pi$- 3$\pi$ and 4$\pi$ mass distributions as well as angular correlations
for 2$(\pi^+\pi^-\pi^0)$ final state at $E_{c.m.} > 1.8$ GeV by contributions of $\omega(782)3\pi$, 
$a_0(980)\rho(770)$ and $\rho(770)4\pi$ intermediate states. Experimental 4$\pi$ mass spectra demonstrate presence of
narrow state. One of the candidate for this state is $f_0(1370)$ with dominant decay into $\rho(770)\rho(770)$ and 
decays into 2$(\pi^+\pi^-)$ and $\pi^+\pi^-2\pi^0$. So, the next step in the study of the 2$(\pi^+\pi^-\pi^0)$
dynamics will be creation of the primary Monte-Carlo generator of the process $e^+e^-\to f_0(1370)(2\pi)_{P-wave}$.

\newpage

\subsection{Automation of the leading order calculations for $e^+e^-\to\;$ 
hadrons}
\addtocontents{toc}{\hspace{2cm}{\sl K.~Kolodziej}\par}

\vspace{5mm}

K.~Ko\l odziej

\vspace{5mm}

\noindent
Institute of Physics, University of Silesia, ul. Uniwersytecka 4, PL-40\,007
Katowice, Poland\\

\vspace{5mm}

After some modifications, {\tt carlomat} \cite{carlomat,carlomat2}, 
a program for automatic computation 
of the leading order (LO) 
cross sections of multiparticle reactions, that was originally dedicated 
mainly to description of the processes of production 
and decay of heavy particles such as top quarks, the Higgs boson, or 
electroweak gauge bosons, can be used to obtain predictions for 
$e^+e^-\to {\rm hadrons}$ in the framework of effective models.
At low energies, the hadronic final states consist mostly of pions, kaons,
or nucleons which can be accompanied by one or more photons, or light fermion 
pairs such as $e^+e^-$, or $\mu^+\mu^-$.
Some effective models which can be useful in this context, including the
scalar electrodynamics (sQED) and the $Wtb$ interaction with operators 
of dimension up to 5, were already implemented in version 2 of 
the program \cite{carlomat2}.

The effective Lagrangian of the $Wtb$ interaction has the following form
\cite{kane}:
\begin{eqnarray}
L_{Wtb}=\frac{g}{\sqrt{2}}\,V_{tb}\left[W^-_{\mu}\bar{b}\,\gamma^{\mu}
\left(f_1^L P_L +f_1^R P_R\right)t 
-\frac{1}{m_W}\partial_{\nu}W^-_{\mu}\bar{b}\,\sigma^{\mu\nu}
  \left(f_2^L P_L +f_2^R P_R\right)t\right]+{\rm h.c.},
\end{eqnarray}
where the couplings $f_{i}^{L}$, $f_{i}^{R}$, $i=1,2$, can be complex in general. 
The electromagnetic (EM) interaction of spin 1/2 nucleons has a similar form:
\begin{eqnarray}
L_{\gamma NN}&=&eA_{\mu}\bar{N}(p')\left[\gamma^{\mu} F_1(Q^2)
+\frac{i}{2m_N}\sigma^{\mu\nu}q_{\nu} F_2(Q^2)\right]N(p).
\end{eqnarray}
The form factors $F_1(Q^2)$ and $F_2(Q^2)$, where $Q^2=-(p-p')^2$, were
adopted from {\tt PHOKARA} \cite{PHOKARA}, thus making possible
Monte Carlo (MC) simulations of processes involving the EM interaction 
of nucleons.

\begin{figure}[htb]
\centerline{
\includegraphics[width=52mm, height=26mm]{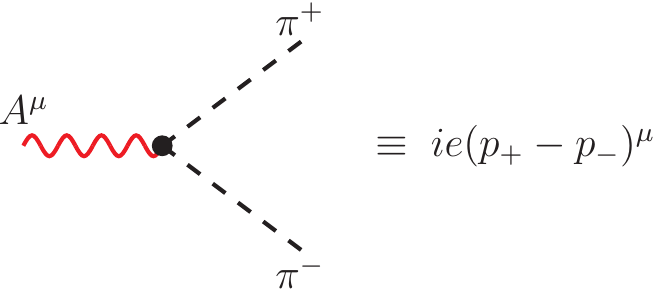}
\hspace*{1cm}
\includegraphics[width=52mm, height=26mm]{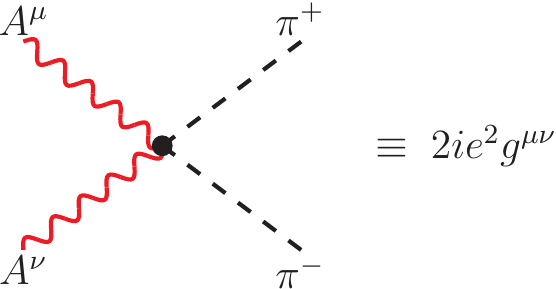}}
\caption{\small Vertices of sQED}
\label{vertsQED}
\end{figure}

At low energies, $\pi^{\pm}$ can be treated as point like particles 
and their EM interaction can be effectively described
in the framework of sQED \cite{fred} the interaction vertices examples
of which are shown in Fig.~\ref{vertsQED}.

\begin{figure}[htb]
\begin{tabular}{ll}
\hspace*{1.cm}
\includegraphics[width=54mm, height=24mm]{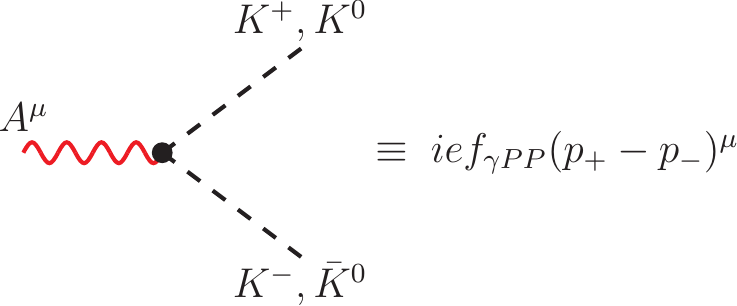}&
\hspace*{2.cm}
\includegraphics[width=54mm, height=24mm]{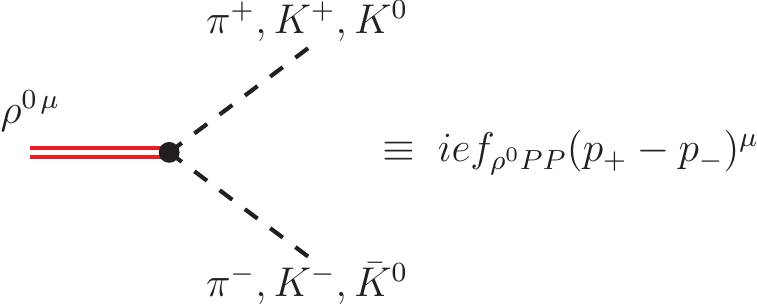}\\[2mm]
\hspace*{1.cm}
\includegraphics[width=54mm, height=24mm]{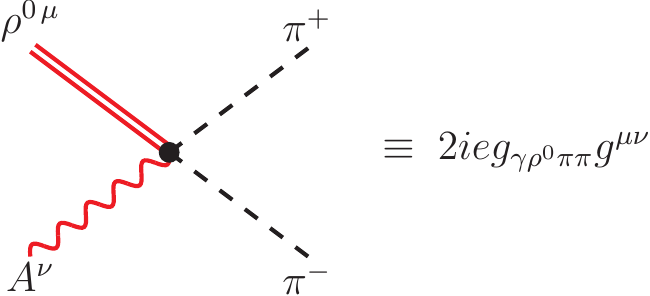}&
\hspace*{2.cm}
\includegraphics[width=54mm, height=24mm]{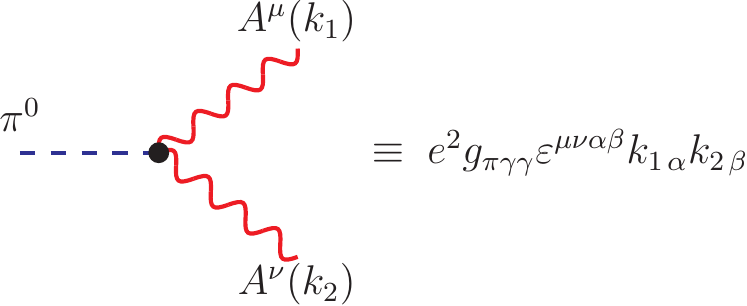}\\[2mm]
\hspace*{1.cm}
\includegraphics[width=54mm, height=24mm]{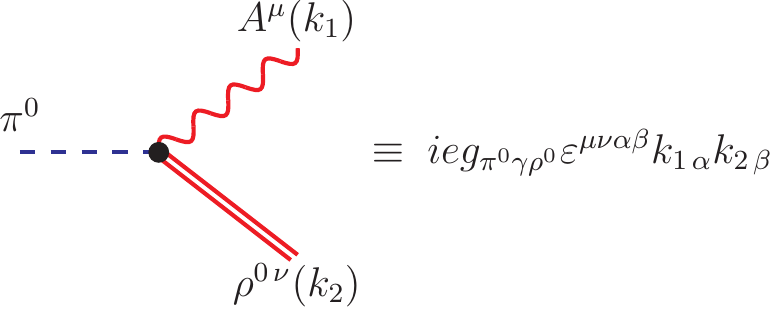}&
\hspace*{2cm}
\includegraphics[width=66mm, height=24mm]{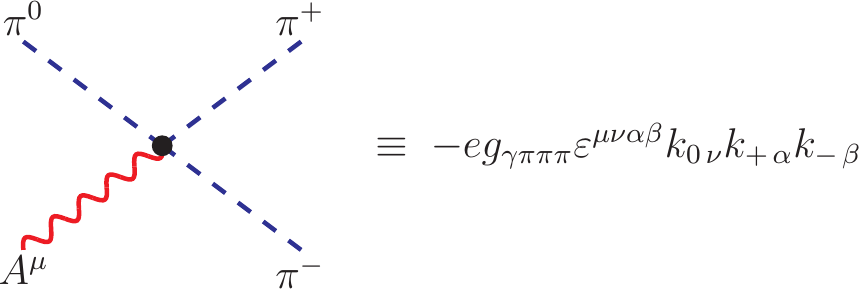}
\end{tabular}
\caption{\small Examples of triple and quartic vertices of RChPT.}
\label{vertRChPT}
\end{figure}

\begin{figure}[!ht]
\centerline{
\includegraphics[width=54mm, height=7.2mm]{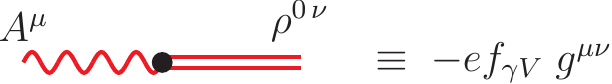}
\hspace*{1cm}
\includegraphics[width=54mm, height=7.2mm]{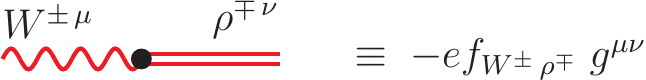}}
\caption{\small Examples of the particle mixing.}
\label{mixing}
\end{figure}

Another step toward better description of $e^+e^-\to {\rm hadrons}$ 
at low energies is the inclusion of the Feynman rules 
of the Resonance Chiral Perturbation Theory 
(RChPT). The interaction vertices and particle mixing terms of RChPT that can be
relevant in this context were provided by Fred Jegerlehner \cite{FJ}. 
Some examples of them are shown in Figs.~\ref{vertRChPT} and \ref{mixing}.
The implementation of the triple and quartic interaction vertices 
was more or less straightforward, as it just
required writing  a few new subroutines for computation of 
the helicity amplitudes involving the Lorentz 
tensors that are different from those of the sQED vertices. The couplings 
$f_{\gamma PP}$, $f_{\rho^0 PP}$, 
$g_{\gamma\rho^0\pi\pi}$, $g_{\pi\gamma\gamma}$, $g_{\pi^0\gamma\rho^0}$
and $g_{\gamma\pi\pi\pi}$ are currently set either to 1 or $e$.
However, implementation of the particle mixing is more challenging, 
because it must be added at the stage, where the topologies of
diagrams which, in {\tt carlomat}, contain only triple and quartic vertices,
are confronted with the Feynman rules.
This required substantial changes in the code generating part of the program.

To illustrate how the program works, consider the process 
$e^+ e^- \to \pi^+ \pi^- \pi^+ \pi^- \gamma$. Taking into account the Feynman 
rules of the standard model and the rules of Figs.~\ref{vertsQED}, 
\ref{vertRChPT} and \ref{mixing}, {\tt carlomat} generates the $U(1)$ gauge 
invariant matrix element, which receives contributions from 903 LO Feynman
diagrams,  together with a dedicated multichannel phase space integration 
routine in just a few seconds. A computation of the total cross 
section, including any number of differential distributions, which is 
performed as the next step, takes several dozen seconds or
several minutes time, dependent on the desired precision of the MC integration.

This project was supported in part with financial 
resources of the Polish National Science Centre (NCN) under grant decision 
No. DEC-2011/03/B/ST6/01615.

\newpage

\subsection{MCGPJ for the processes $e^{+}e^{-}{\rightarrow}hadrons$ for experiments with CMD-3 detector at the VEPP-2000 collider}
\addtocontents{toc}{\hspace{2cm}{\sl G.V.~Fedotovich}\par}

\vspace{5mm}

G.V.~Fedotovich, V.L.~Ivanov, D.N.~Shemyakin

\vspace{5mm}

\noindent
Budker Institute of Nuclear Physics, SB RAS, Novosibirsk, 630090, Russia\\
Novosibirsk State University, Novosibirsk, 630090, Russia
\vspace{5mm}

The hadronic contribution to $(g-2)/2$ of muon, coming from VEPP-2000 energy range, is about $92\%$. One of the aims of the experiments with CMD-3 detector is to measure the main multihadrons cross sections (MHCS) with systematic uncertainty smaller than $3\%$ (${\sim}0.15~\rm ppm$ in $(g-2)/2$). The systematic accuracy of the cross section for the channel $e^{+}e^{-}{\rightarrow}\pi^{+}\pi^{-}$ at least better than $0.5\%$ is required, as it is seen from the ``bench'' estimation. It is well known that the hadronic contribution to anomalous magnetic moment of muon (AMM) is about $60~\rm ppm$: $0.005{\times}60~\rm ppm = 0.3~\rm ppm$. The aim of the new FNAL and JPARC experiments to measure the $(g-2)/2$ of muon is to improve the previous BNL result by a factor of 4 and to achieve the accuracy of ${\sim}0.15~\rm ppm$ - good test of the SM. It is obvious that the systematic uncertainly of radiative corrections (RC) for the MHCS should be smaller than $1-2\%$.

Previous experience of the studying of $3\pi$ channel at CMD-2 confirms, that final state radiation (FSR) contributes to the cross section (CS) at the level of $0.4\%$ ~\cite{1}. Obviously, that for the multihadron channels the contribution of FSR to CS will be smaller than $0.4\%$. In the scale of $1-2\%$ for expected systematic accuracy we can neglect by FSR and consider photon jets radiation in collinear regions only. After many discussions with our experts in Dubna (JINR) we chose the following strategy, which will be described using the channels $e^{+}e^{-}{\rightarrow}K^{+}K^{-}\pi^{+}\pi^{-}$ and $e^{+}e^{-}{\rightarrow}K^{+}K^{-}\eta$ as the examples. 

To select the clean signal events at first step four charged particles with zero net charge are selected using information from the drift chamber (DC). At the second step after the procedure of separation of kaons and pions (using $dE/dx$ information ~\cite{2}) we calculate the values: ${\Delta}E=E_{1}+E_{2}+E_{3}+E_{4}-2E_{\rm beam}$ and $|\vec{p}|=|\vec{p}_{1}+\vec{p}_{2}+\vec{p}_{3}+\vec{p}_{4}|$. Two dimensional plot ${\Delta}E$ vs $|\vec{p}|$ for selected events is shown in Fig.1. The events of the signal process are inside the rectangle.  The analysis of these events revealed that at least four intermediate states exist: $K^{*}\bar{K}^{*}$, $\phi\pi^{+}\pi^{-}$, $K^{+}K^{-}\rho$, $K^{*}(892)K\pi$. The simplest model was chosen to describe (more or less) correctly the experimental angular and momentum distributions. In Figures 2-5 the results of simulation vs experiment are presented. To increase the statistics we combine the data collected at several energy points. Red points correspond to experiment, black points - to simulation according to "realistic" model mentioned above, blue points - to simulation according to the $K^{+}K^{-}\pi^{+}\pi^{-}$ phase space. When the $K^{+}K^{-}\pi^{+}\pi^{-}$ events are selected and the dynamics of their production is defined we are able to calculate the detection efficiency and to determine the visible CS. In order to calculate the RC $(1+\delta_{\rm rad}(s))$ we calculate the following integral with Structure Functions $D(x,s)$ ~\cite{3}, which describe photon jets radiation in the collinear regions:

\begin{eqnarray}
\sigma_{\rm vis}(s)=\int\limits_{0}^{1}dx_{1}\int\limits_{0}^{1}dx_{2}\,D(x_{1},s)D(x_{2},s)\sigma_{\rm born}(s(1-x_{1})(1-x_{2}))=(1+\delta_{\rm rad}(s))\sigma_{\rm born}(s). \nonumber
\end{eqnarray}

At first iteration in the integral we use born CS, measured by BABAR (if there were not previous measurements of the born CS, then instead of the latter the visible CS should be used). This procedure is repeated for several iterations while RC does not become stable inside a corridor of ${\sim}0.3\%$. The results of such calculation are plotted in Fig.6. The uncertainties of the RC are caused by the uncertainty of the form of the CS.

\begin{figure}[tbh]
\begin{minipage}[t]{0.48\textwidth}
\centerline{\includegraphics[width=1.0\textwidth]{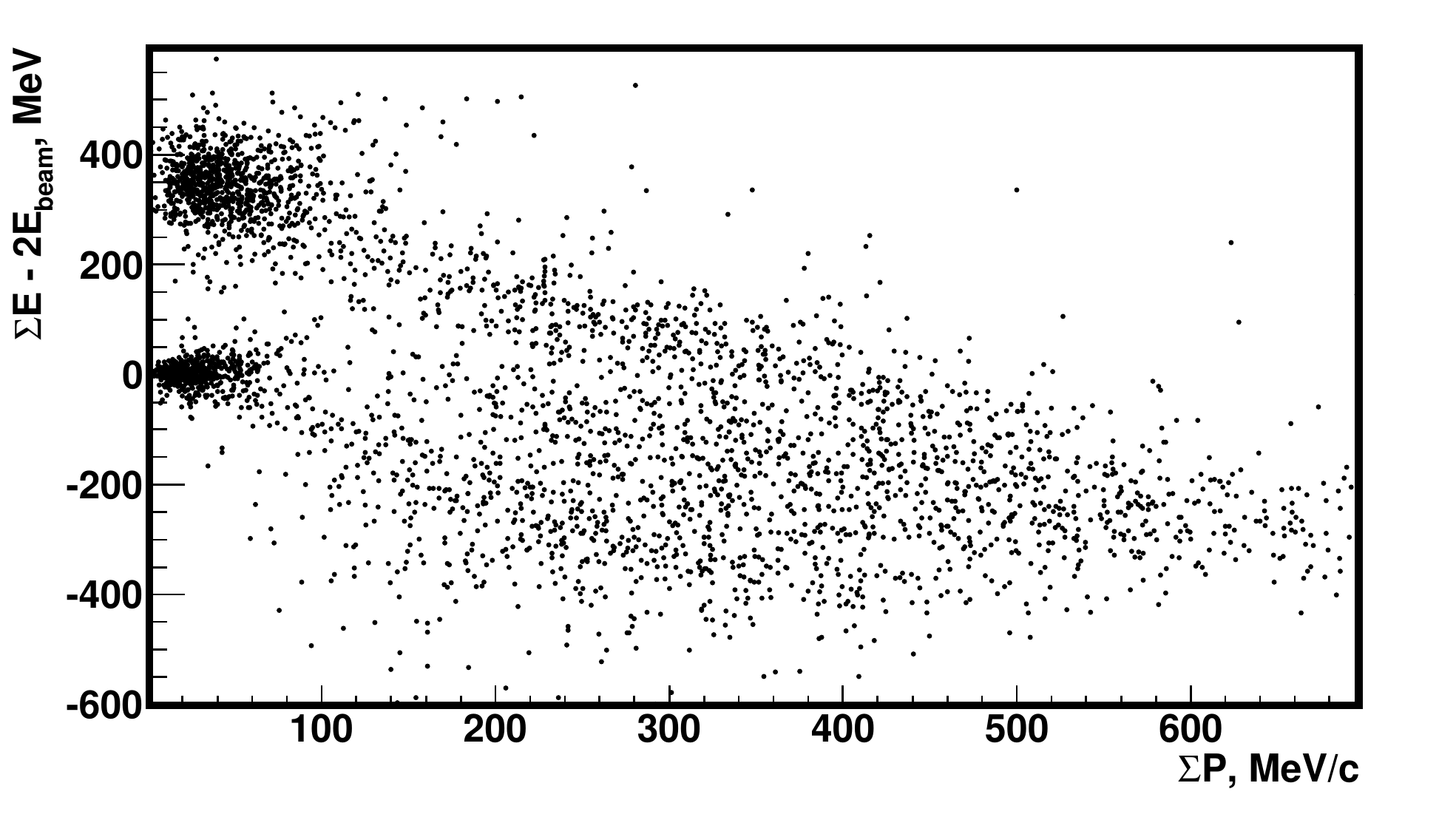}}
\caption{Two dimensional plot ${\Delta}E$ vs $|\vec{p}|$ for selected events. The $K^{+}K^{-}\pi^{+}\pi^{-}$ events are inside the rectangle.}
\end{minipage}\hfill\hfill
\begin{minipage}[t]{0.48\textwidth}
\centerline{\includegraphics[width=1.0\textwidth]{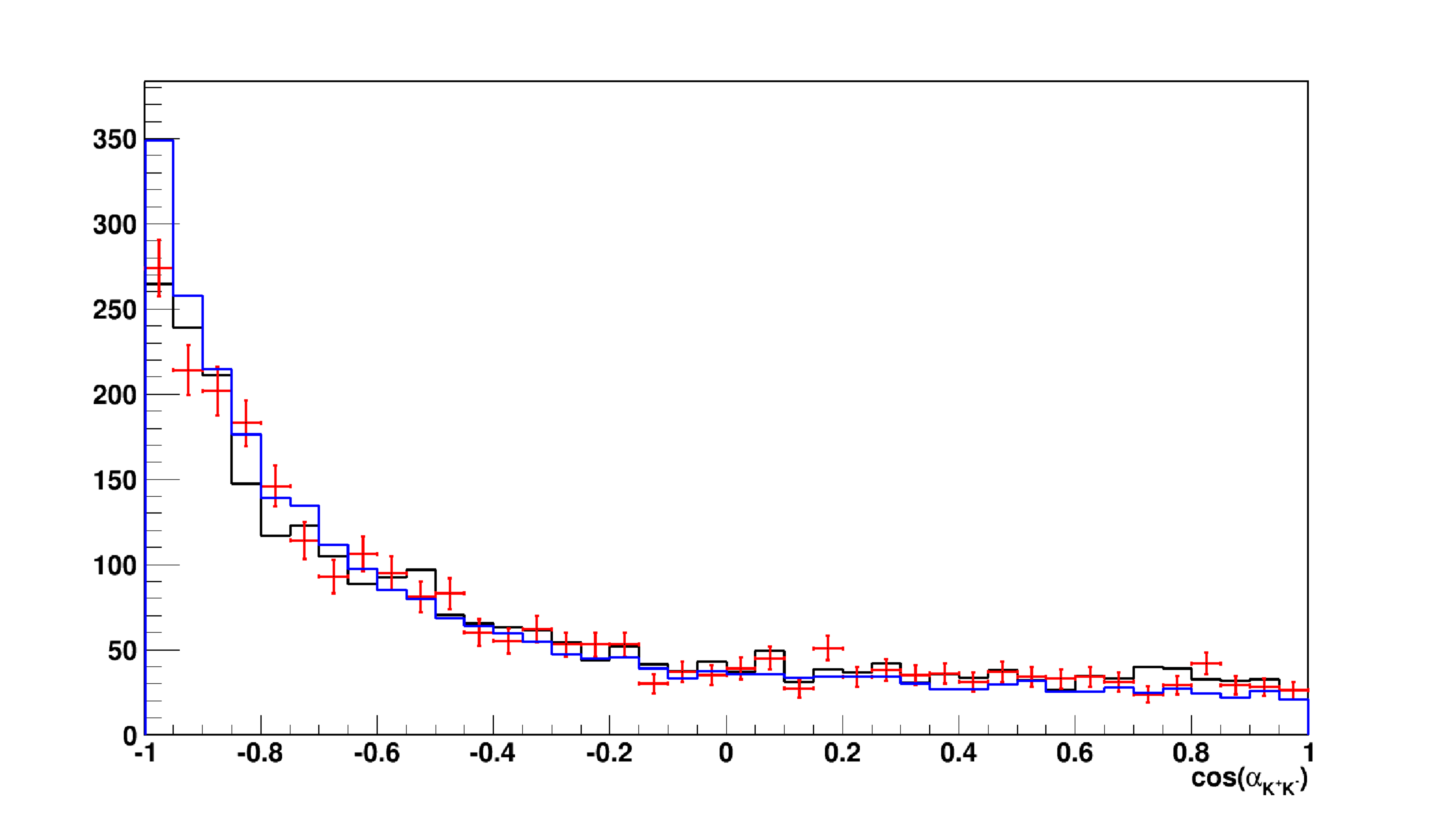}}
\caption{The distribution of the cosine of angle between $K^{+}$ and $K^{-}$.}
\end{minipage}
\end{figure}

\begin{figure}[tbh]
\begin{minipage}[t]{0.48\textwidth}
\centerline{\includegraphics[width=1.0\textwidth]{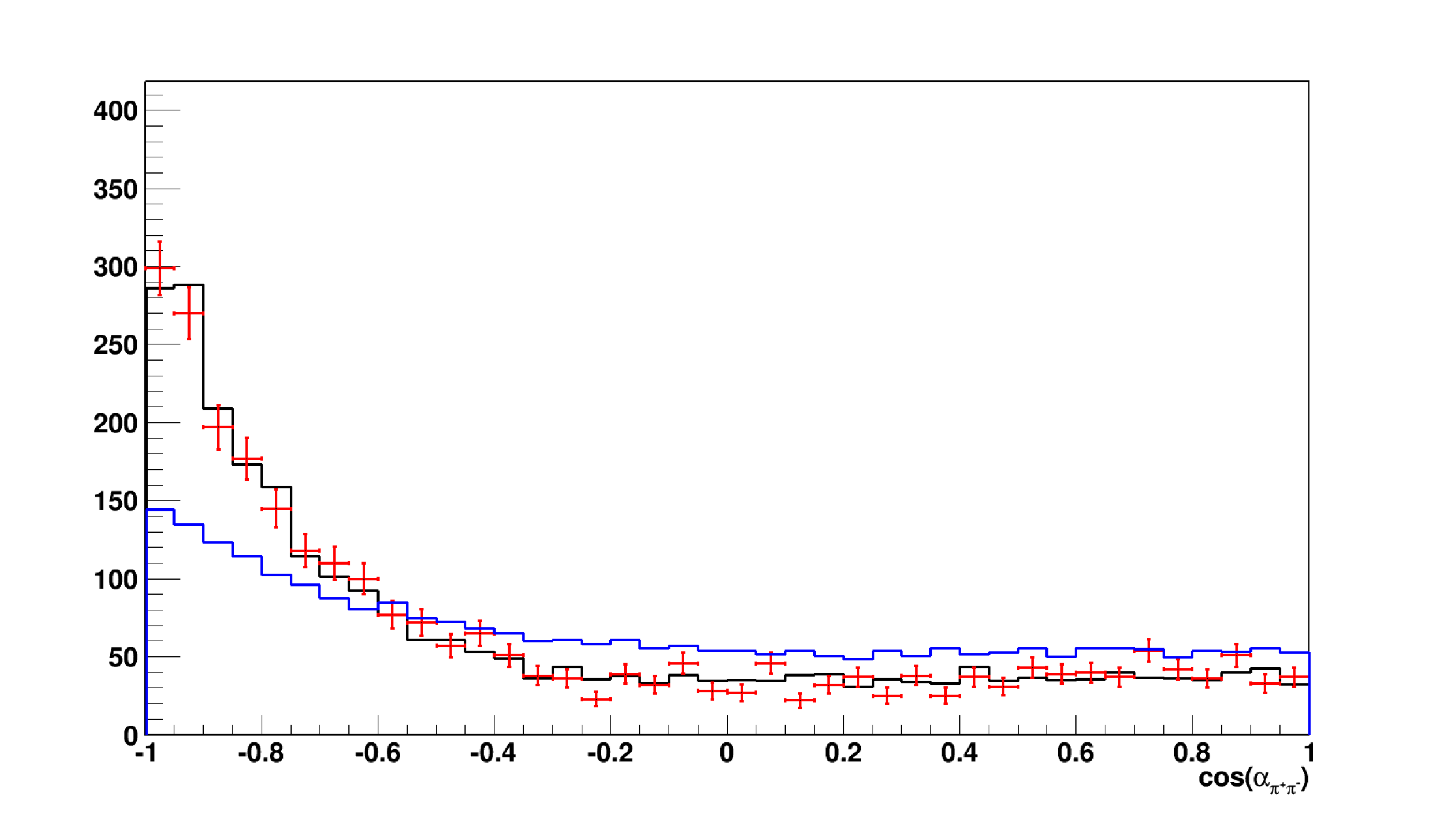}}
\caption{The distribution of the cosine of angle between $\pi^{+}$ and $\pi^{-}$.}
\end{minipage}\hfill\hfill
\begin{minipage}[t]{0.48\textwidth}
\centerline{\includegraphics[width=1.0\textwidth]{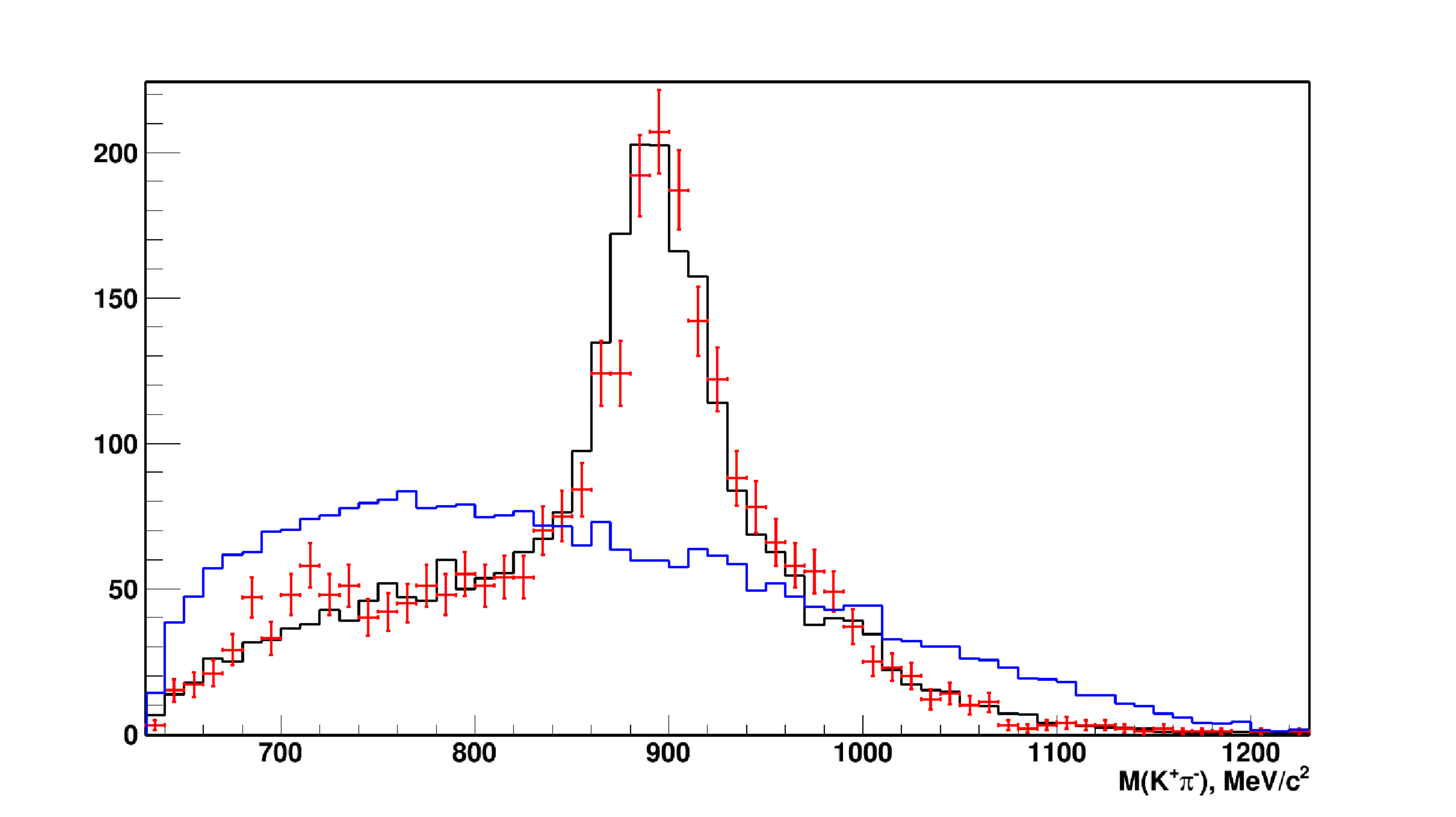}}
\caption{The distribution of the invariant mass of $K^{-}\pi^{+}$, $K^{+}\pi^{-}$.}
\end{minipage}
\end{figure}


This final state is almost completely produced by the $\phi(1680){\rightarrow}\phi\eta$ mechanism. In our analysis we do not reconstruct the $\eta$ from its decay products and do not use the information from calorimeters. This approach allows us to use all the modes of $\eta$ decay and thus enlarges the statistics, but complicates background subtraction. The main sources of background here are $K^{+}K^{-}\pi^{+}\pi^{-}$ and $K^{+}K^{-}\pi^{0}\pi^{0}$ (especially their ${\phi}f_{0}(600)$ intermediate mechanism).

\begin{figure}
\begin{minipage}{0.47\textwidth}
\centerline{\includegraphics[width=1.0\textwidth]{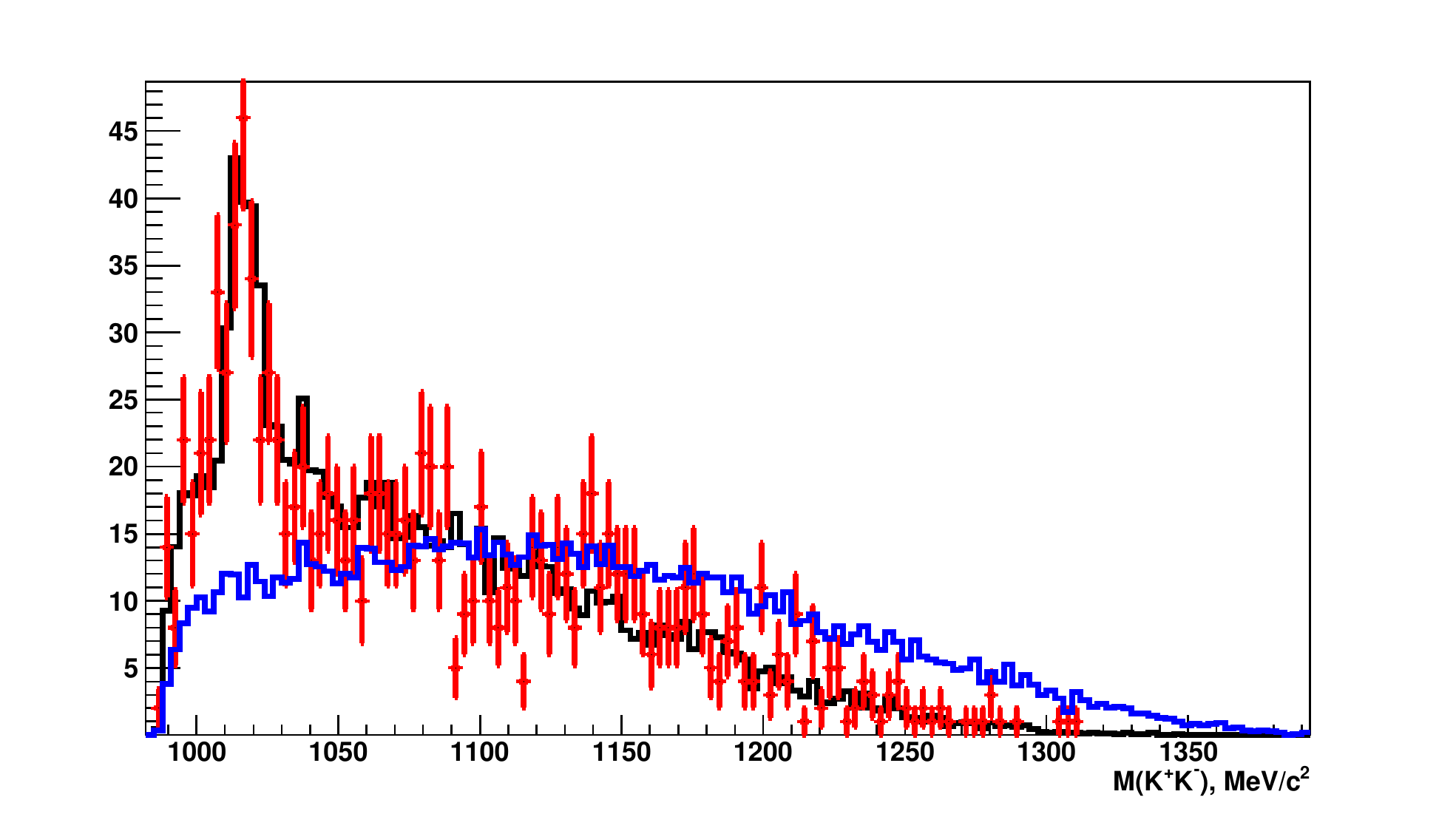}}
\caption{The distribution of the invariant mass of $K^{+}K^{-}$.}
\end{minipage}\hfill\hfill
\begin{minipage}{0.47\textwidth}
\centerline{\includegraphics[width=1.0\textwidth]{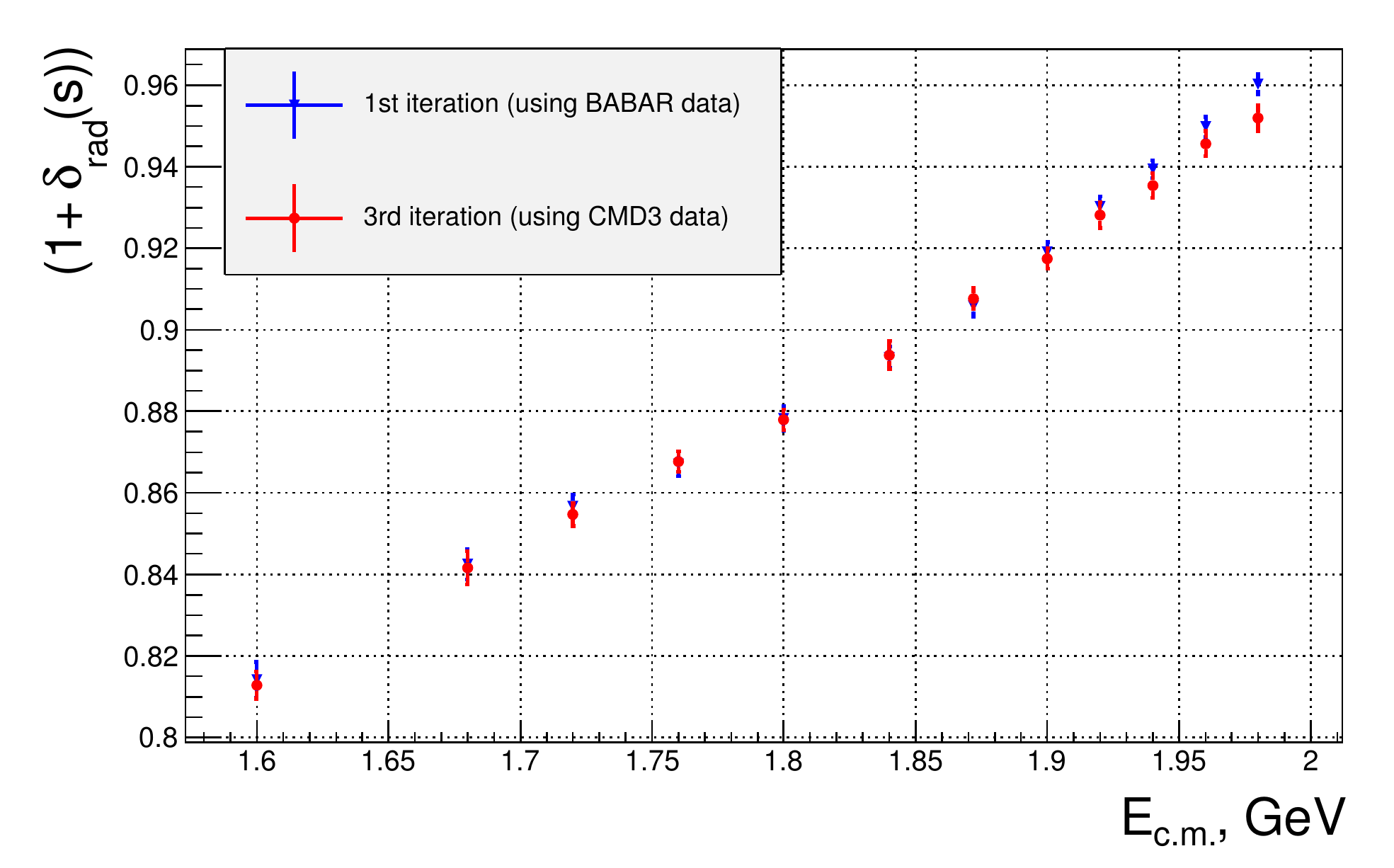}}
\caption{RC for $e^{+}e^{-}{\rightarrow}K^{+}K^{-}\pi^{+}\pi^{-}$: blue - 1st iteration on the base of BABAR born CS, red - 3rd iteration on the base of CMD-3 CS.}
\end{minipage}
\begin{minipage}{0.47\textwidth}
\centerline{\includegraphics[width=1.0\textwidth]{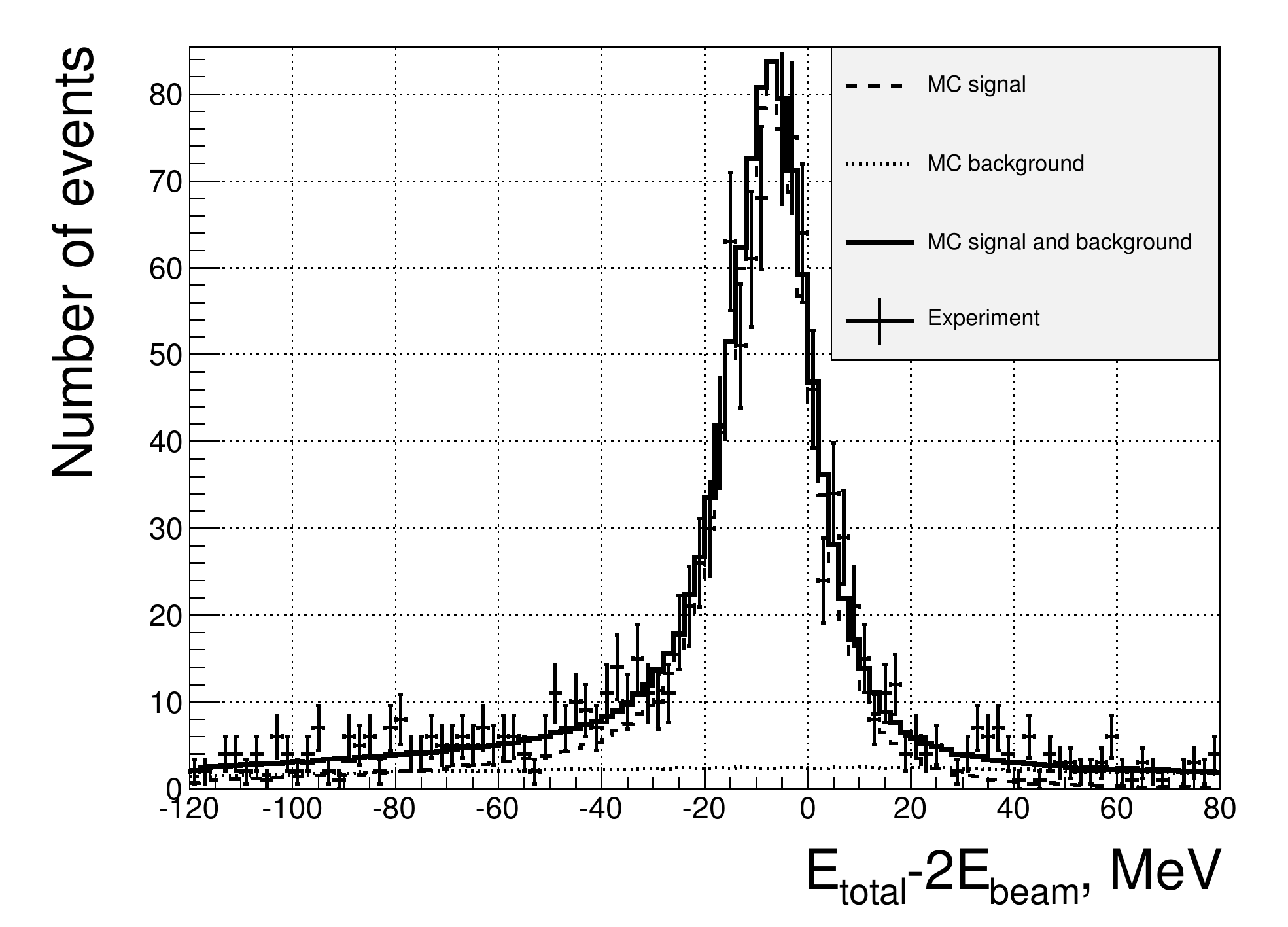}}
\caption{Results of separation of the signal and background: distribution of the parameter $E_{\rm total}-2E_{\rm beam}$ (all energy points are combined) for 1) experiment (points with error bars) 2) signal events (open histogram with dashed line) 3) background events (open histogram with dotted line) 4) sum of signal and background events (open histogram with solid line).}
\end{minipage}\hfill\hfill
\begin{minipage}{0.47\textwidth}
\centerline{\includegraphics[width=1.0\textwidth]{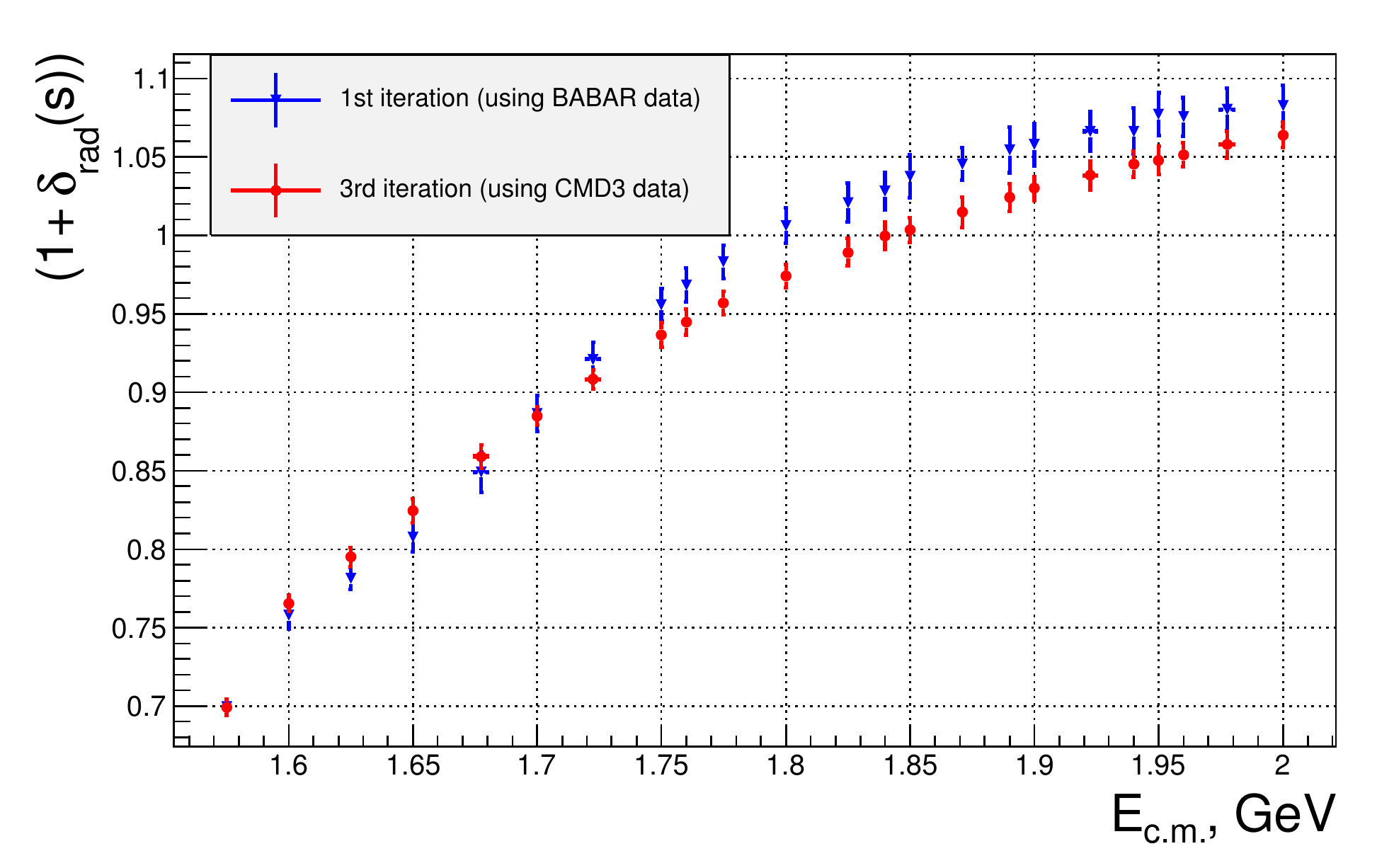}}
\caption{RC for $e^{+}e^{-}{\rightarrow}K^{+}K^{-}\eta$: blue - 1st iteration on the base of BABAR born CS, red - 3rd iteration on the base of CMD-3 CS.}
\end{minipage}
\end{figure}

Using the information from DC we calculate the parameter $E_{\rm total}-2E_{\rm beam}$:

\begin{eqnarray}
E_{\rm total}-2E_{\rm beam}=\sqrt{\vec{p}_{K^{+}}^{2}+m^{2}_{K}}+\sqrt{\vec{p}_{K^{-}}^{2}+m^{2}_{K}}+\sqrt{(-\vec{p}_{K^{+}}-\vec{p}_{K^{-}})^{2}+m^{2}_{\eta}}-2{\cdot}E_{\rm beam}, \nonumber
\end{eqnarray}
\noindent
which represents the total energy of the final particles minus twice beam energy in assumption that the missing particle is the $\eta$-meson. The distribution of this parameter for the $K^{+}K^{-}\eta$ has a peak around zero for any energy point and is used for defining the number of selected signal events $N^{\rm exp}_{K^{+}K^{-}\eta}$. The shape of the distribution of this parameter for background is determined from MC. The results of the fitting of the experimental distributions are summarized in the Fig.7.

Having gotten the $N^{\rm exp}_{K^{+}K^{-}\eta}$ and the detection efficiency we calculate the visible CS. The calculation of the RC is performed according the procedure, described above and shown in Fig.8.

In summary, we pointed out that until the precision of ${\sim}1-2\%$ the multihadron cross sections can be studied without taking into account the FSR, i.e. considering photon jets radiation in collinear regions only. As the examples of such approach we have considered the calculation of RC for the processes $e^{+}e^{-}{\rightarrow}K^{+}K^{-}\pi^{+}\pi^{-}$ and $e^{+}e^{-}{\rightarrow}K^{+}K^{-}\eta$, which are under study at CMD-3.

\newpage

\subsection{$\chi_{c1}$ and $\chi_{c2}$ production at $e^+ e^-$ colliders - 
preliminary results}
\addtocontents{toc}{\hspace{2cm}{\sl S.~Tracz}\par}

\vspace{5mm}

\underline{S.~Tracz}, H.~Czy\.z, P.~Kisza

\vspace{5mm}

\noindent
Institute of Physics, University of Silesia, Katowice, Poland
\vspace{5mm}

With the improving luminosity of $e^+ e^-$ colliders, the search for a 
  production of $0^{++}$, $1^{++}$ and $2^{++}$ states become possible. 
The production of these states goes through two virtual photons. 
 The amplitude describing creation of the $0^{++}$ state through reaction
  $e^+e^-\to \chi^*_{c0}\to \cdots$,
  going through loop diagram \cite{Kuhn:1979bb},
 is proportional to electron mass and thus highly suppressed. All $\chi_c$
 states 
 can be however produced through $e^+e^-\to e^+e^-\chi^*_{ci}(\to \cdots)$,
  $i=0,1,2$, reaction.

  Measurements of the cross sections of the reaction
 $e^+e^-\to \chi^*_{c1,c2}\to \cdots$
  will allow to measure the electronic widths ($\Gamma^{\chi_{c1,c2}}_{e^+e^-}$) of
 the $\chi_{c1}$ and $\chi_{c2}$ resonances. Combined with measurements
 of the differential cross section of the reactions 
 $e^+e^-\to e^+e^-\chi^*_{ci}(\to \cdots)$
 they will allow for detailed tests of  models  describing these 
 charmonium bound states. 

 Expected range of $\Gamma^{\chi_{c1,c2}}_{e^+e^-}$ have been calculated inside 
two models, quarkonium model and vector dominance model already
  in \cite{Kuhn:1979bb}.
 Within the quarkonium model the amplitudes describing coupling
  of two virtual photons to $J^{++}$ states 
 depend on binding energy and the derivative of the wave 
 function at the origin. This model predicts also that only some
 of the allowed amplitudes contribute.
 For $\chi_{c0}$ from two allowed amplitudes only one contributes. 
For  $\chi_{c1}$ from three independent amplitudes one gives contribution.
 While for $\chi_{c2}$ from five possible amplitudes only one gives
  contribution. 

\begin{figure}[t]
\begin{center}
\includegraphics[trim=4cm 25cm 0cm 0cm, clip=true, totalheight=0.25\textheight]{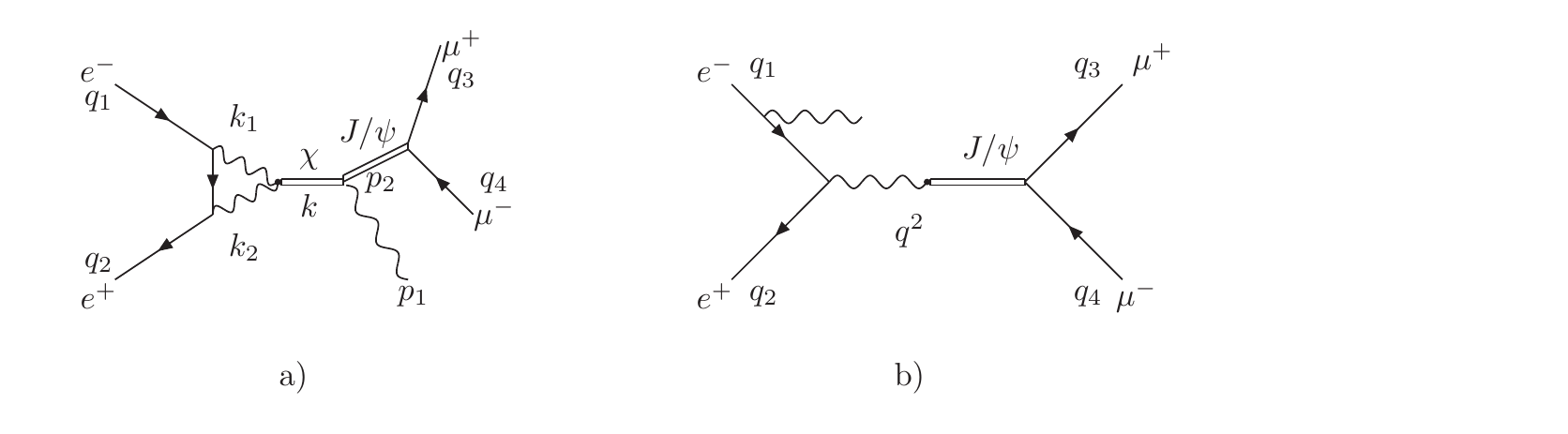}
\caption{
a) Diagram for process $e^+e^- \rightarrow \mu^+\mu^-$ with $\chi_{c1,2}$ production b) Diagram for radiative return process.
 \label{fig1}
}
\end{center}
\end{figure}

 For production of $\chi_{c0,1,2}$ states one can concentrate on a selected 
 final state which is easy to be observed experimentally,
 mainly the decay of $\chi_{c0,1,2}$ into $J/\psi \gamma$,
  where $J/\psi$ subsequently decays into pair of muons.
 For the   $e^+e^-\to \chi^*_{c1,c2}\to J/\psi (\to \mu^+\mu^-) \gamma $
  the process has to be considered together with radiative return process,
 which is a non-reducible background (see Figure \ref{fig1} for diagrams).
  Furthermore one should take into account the unknown 
  relative phase of the signal
  and the background amplitudes, which could give
  an interesting interference pattern. 
 The binding energies and the derivative of the wave 
 functions at the origin for $\chi_{c0,1,2}$ can be extracted
 from known \cite{Agashe:2014kda} values of 
 $\Gamma(\chi_{c0,1,2}\rightarrow \gamma \gamma)$
 and $\Gamma(\chi_{c0,1,2,3} \rightarrow J/\psi \gamma)$
Assuming that the binding energies are different for each state 
 and the derivative of the wave 
 functions are equal,
 we have performed four fits.
  First fit was done to the data for $\chi_{c1}$ and $\chi_{c2}$,
  second to the data for $\chi_{c0}$ and $\chi_{c1}$, 
  third to the data for $\chi_{c0}$ and $\chi_{c2}$
  and fourth one to all data. The obtained results show 
 that it is impossible to fit simultaneously 
 the data for the states $\chi_{c0}$ and $\chi_{c2}$. 
 Thus the quarkonium model is not able to accommodate these data,
 even if the discrepancy is not dramatic.
  In the case of the global fit only width 
 $\Gamma(\chi_{c0}\rightarrow \gamma \gamma)$ does not fit well. 
 One has to remember however that the model is non-relativistic, 
 while obtained binding energies and hence velocities of quarks are large.
  Using the fitted parameters we have made predictions  of  electronic widths.
 They are not greater than $0.1eV$, which is at the limit 
 of BES-III sensitivity.

 To examine a possibility of studies of $\chi_c-\gamma^*-\gamma^*$
  amplitudes at meson factories, we have 
  calculated the cross section of the reaction
  $e^+e^-\to e^+e^-\chi^*_{ci} \to J/\psi (\to \mu^+\mu^-) \gamma$
  within the same model. 
 We have assumed integrated luminosities $L_{int}=10fb^{-1}$ 
 at energy 4.23 GeV(BES-III), $L_{int}=530 fb^{-1}$ (BaBar),
  $L_{int}=1000 fb^{-1}$ (BELLE)
  and  $L_{int}=50ab^{-1}$ (BELLE-2). In the last three we have used energy 
 10.56 GeV. For BES-III energy and luminosity the event rates are too small
 to be observed. The biggest expected rate is for $\chi_{c2}$ with 20 expected
  events if the final electron and positron are not tagged.
 For BaBar, BELLE and BELLE-2 the expected event rates with no electron-positron
 tagging are 1700,3300,160000 for $\chi_{c0}$, 19000,36000,1800000
  for $\chi_{c1}$
 and 26000,50000,2500000 for $\chi_{c2}$. 
  For BELLE-2, even if one tags both electron and positron within
 the angular range  $20^o-160^o$, the expected event rates are
 big enough to be observed and are equal to 7500, 400000,100000 for
  $\chi_{c0,c1,c2}$ respectively.
  The stricking difference between $\chi_{c0,c2}$ and $\chi_{c1}$ 
 seen here comes from the fact that for $\chi_{c1}$ the contribution
  from real photons is equal to zero and thus for small
  photons virtualities the amplitude is small.

The modes of production of $\chi_{c1,2}$ with the subsequent decay described
  above have been implemented in PHOKHARA Monte Carlo generator and
  the $\chi_{c0,c1,c2}$ production and decay mode
  was added in EKHARA Monte Carlo generator. 
 Preliminary results shown above indicate that $\chi_{c0,1,2}-\gamma^*-\gamma^*$
 amplitudes can be studied in existing or near future experiments.
  More detailed analysis will be presented in \cite{c1}.
\vspace{10mm}


Work 
supported in part by
the Polish National Science Centre, grant number DEC-2012/07/B/ST2/03867 and
German Research Foundation DFG under
Contract No. Collaborative Research Center CRC-1044. 
Szymon Tracz and Patrycja Kisza are supported by the Forszt project co-financed 
by EU from the European Social Fund.

\newpage

\subsection{Nucleon form factors in PHOKHARA}
\addtocontents{toc}{\hspace{2cm}{\sl  H.~Czy\.z}\par}

\vspace{5mm}
\underline{H.~Czy\.z}, S.~Tracz

\vspace{5mm}

\noindent
 Institute of Physics, University of Silesia,
PL-40007 Katowice, Poland

\vspace{5mm}

The electromagnetic  nucleon form factors were studied
  from the conception of particle 
 physics
  \cite{Stern,Rosenbluth:1950yq,Mcallister:1956ng,Hofstadter:1956qs}. 
 and yet a lot has to be done to built a model which meets 
  requirements in the era of precision hadronic physics.
 The not expected developments in this field show that
  we still have much to learn.
 The measurements of the ratio of the electric and magnetic form factors of the 
 proton  with two different methods were giving
  different results \cite{Arrington:2003df}.
 The two-photon exchange radiative corrections explaining 
 to large extent this difference (see \cite{Carlson:2007sp} for review)
 turned out to be  unexpectedly large. Moreover their modelling 
 goes beyond the nucleon form factor modelling adding to the complexity 
 of the problem. In addition, usually the models are built separately
 in the space-like (see \cite{Perdrisat:2006hj} for review)
  or time-like regions (see \cite{Denig:2012by} for review). 
  As one expects that each of the form factors is a unique analytic
 function valid in both space-like and time-like regions this
  attitude has to be changed and a model describing both regions
 has to be constructed. A step towards such a model was done
 in \cite{Czyz:2014sha}. The model describes well data
 from both space-like and time-like regions. The form factors are normalised
  properly at zero invariant mass and, by construction,
  have correct behaviour \cite{Lepage:1980fj} at large invariant masses.
 Yet the model is far from being satisfactory and further studies
 of photon-nucleon interactions are necessary. It is also clear that
 the progress can be achieved only through close collaboration of experimental
 and theory groups. Careful studies of charge and/or
   forward-backward asymmetries 
 in $e^+e^-\to \bar p p$, $e^+e^-\to \bar p p\gamma$ processes together
  with
 angular distributions in $e^-p\to e^-p$ and $e^+p\to e^+p$ processes
  should allow for
  disentangling of the two-photon exchange contributions from the
  form factors. 

   Model testing is simplified if it is implemented into
  a Monte Carlo event generator. Such a generator serves also 
  for other purposes like calculations of acceptance and/or 
  efficiency corrections. For the radiative return (called also ISR)
  method such a tool was developed some time ago \cite{Rodrigo:2001kf}. 
  Nucleon final states were implemented in it already in \cite{Czyz:2004ua}
  and the nucleon form factors were updated in \cite{Czyz:2014sha}, where
  also the modelling of the final state radiative corrections was addressed.
  Recently \cite{Czyz:2013xga} also a possibility of generation of the
  process $e^+e^-\to$~hadrons, useful for scan experiments, was added.


Work 
supported in part by
the Polish National Science Centre, grant number DEC-2012/07/B/ST2/03867 and
German Research Foundation DFG under
Contract No. Collaborative Research Center CRC-1044. 
Szymon Tracz is supported by the Forszt project co-financed 
by EU from the European Social Fund.

\newpage

\subsection{ Current status of Monte Carlo generator Tauola}
\addtocontents{toc}{\hspace{2cm}{\sl O.~Shekhovtsova}\par}

\vspace{5mm}

O.~Shekhovtsova

\vspace{5mm}
\noindent
 Institute of Nuclear Physics PAN, ul. Radzikowskiego 152, 31-342 Krakow, Poland \\
           Kharkov Institute of Physics and Technology,  Akademicheskaya,1, 61108 Kharkov, Ukraine \\

\vspace{5mm}

The Monte Carlo generator TAUOLA \cite{Jadach:1993hs}, which is used to simulate tau lepton decays,
 is a computing project with a rather long history. It started in the years 80's and is under development still now. The main problem in the theoretical description of the hadronic decay modes of the tau lepton is a lack of a theory coming from the first principle in the energy region from the threshold till the tau mass. 
The hadronic currents in the first version of TAUOLA as well as the subsequent internal versions of the code used by both Aleph and Cleo collaborations were based on the Vector Meson Dominance approach. However, that approach is able to reproduce only the leading-order results of Chiral Perturbation Theory (ChPT). As was shown in the case of the $K^+K^-\pi^-$ mode this approach spoils the Wess-Zumino anomaly normalization, that appears at $O(p^4)$ in ChPT \cite{Coan:2004ep}. An alternative approach is to include the lightest resonances as
active degrees of freedom in the theory. This can be done by adding resonance
fields to the ChPT Lagrangian, without any dynamical assumption,  leading to Resonance Chiral Lagrangian approach \cite{Ecker:1988te, Ecker:1989yg}. The RChL approach succeeds in reproducing low energy
results, predicted by ChPT, at least at the next-to-leading order, and also complies with QCD high energy constraints. 

 Up to now the RChL currents for the main two-meson (final states with two pion, pion-kaon, two kaons) and three-pseudoscalar (three pion, two kaon-one pion) decay modes have been installed into TAUOLA. This set covers more than $88\%$ of the tau lepton hadronic decay width. The implementation of the currents, the related technical tests as well as the necessary theoretical concepts are documented in \cite{Shekhovtsova:2012ra}.

To get numerical values of the model parameters one has to fit the theoretically predicted spectra to those measured in experiments. We started with the $\tau^-\to\pi^-\pi^-\pi^+\nu_\tau$ decay.  The first fit to the preliminary BaBar data  \cite{Nugent:2013hxa} allowed to extract numerical values of the model parameters and demonstrated satisfactory agreement with the three pion invariant mass spectrum and a mismatch in the low energy tail of the two pion invariant mass distributions \cite{Shekhovtsova:2013rb}. Improvement has been achieved by adding the $\sigma$ resonance contribution in the hadronic current \cite{Nugent:2013ij}. To fit the data we used the MINUIT package through the ROOT framework and the fit result is presented in Fig. 1 of \cite{Nugent:2013ij}. For the numerical values of the
model parameters see Table 1 in  \cite{Nugent:2013ij}. The goodness of the fit is quantified by $\chi^2/{\rm ndf}=6658/401$. To compare with the previous result \cite{Shekhovtsova:2013rb}
 we have estimated the $\chi^2$ value using the combined statistical and systematic uncertainties and we find $\chi^2/{\rm ndf} = 910/401$ that is eight times better than in  \cite{Shekhovtsova:2013rb}.

The following tests have been done to validate our fitting procedure: asserting statistical errors and correlation coefficients between the model parameters, verifying that the obtained result is a global minimum and performing studies of the systematical errors. A detailed description as well as the results of the tests are presented in  \cite{Nugent:2013ij}.

The fitting procedure for the three pion mode has been generalized to an arbitrary three meson final state. Thus the new fitting framework allows to perform fits for arbitrary tau decay mode, using either Fortran or C++ code. To test it, we have first reproduced the results for the three pion decay mode described above and then fitted the $K^+K^-\pi^-$ currents to the BaBar preliminary data \cite{Nugent:2013hxa}. For the moment we have used
some simplifications for the for  $K^+K^-\pi^-$  current which will be removed in the final version of the fitting
strategy. In addition, fits for    $\tau^-\to\pi^-\pi^0\nu_{\tau}$ decay  mode has been added allowing to fit the RChL two pion form
factor to the Belle parameterization. Next step will be to study the stability of the generalized fitting procedure
and to include the experimental errors.

Another important task concerned the C++ interface for the decay channels of TAUOLA generator. The interface allows to add, substitute or modify TAUOLA decay modes using either C++ or FORTRAN
code. The interface is ready to be used in Tauola++ project (http://tauolapp.web.cern.ch/tauolapp/) as well as
in the FORTRAN environments, where TAUOLA is still being used. Validation of all BaBar currents introduced
along with the C++ environment has been finished and extensive testing of the environment has been
performed.

This research was supported in part by Foundation of Polish Science
grant POMOST/2013-7/12, that is co-financed from European Union, Regional
Development
Fund and from funds of Polish National Science
Centre under decisions  DEC-2011/03/B/ST2/00107.

\newpage

\section{List of participants}

\begin{flushleft}
\begin{itemize}
\item Antonio Anastasi, University of Messina and LNF, {\tt antonioanastasi89@gmail.com}
\item Marcin Berlowski, National Center for Nuclear Research, {\tt Marcin.Berlowski@fuw.edu.pl}
\item Bo Cao, University of Uppsala, {\tt bo.cao@physics.uu.se}
\item Carlo M. Carloni Calame, University of Pavia, {\tt carlo.carloni.calame@pv.infn.it}
\item Henryk Czy\.z, University of Silesia, {\tt henryk.czyz@us.edu.pl}
\item Achim Denig, Universt\"at Mainz, {\tt denig@kph.uni-mainz.de }
\item Gennadiy V. Fedotovich, BINP and NSU, {\tt G.V.Fedotovich@inp.nsk.su}
\item Martin Hoferichter, University of Bern, {\tt hoferichter@itp.unibe.ch}
\item Karol Ko\l odziej, University of Silesia, {\tt karol.kolodziej@us.edu.pl}
\item Andrzej Kupsc, University of Uppsala, {\tt Andrzej.Kupsc@physics.uu.se}
\item Peter Lukin, BINP and NSU, {\tt P.A.Lukin@inp.nsk.su}
\item Pere Masjuan, Universt\"at Mainz, {\tt masjuan@kph.uni-mainz.de}
\item Dario Moricciani, INFN -  Roma Tor Vergata, {\tt moricciani@roma2.infn.it}
\item Massimo Passera, INFN Padova, {\tt passera@pd.infn.it}
\item Olga Shekhovtsova, IFJ PAN, {\tt Olga.Shekhovtsova@lnf.infn.it}
\item Szymon Tracz, University of Silesia, {\tt szymon885@gazeta.pl}
\item Graziano Venanzoni, LNF, {\tt Graziano.Venanzoni@lnf.infn.it}
\item Ping Wang, Institute of High-energy Physics, Beijing, {\tt wangp@ihep.ac.cn}
\end{itemize}
\end{flushleft}

\end{document}